\documentclass[11pt]{article}
\usepackage{
amsmath,amsfonts,amssymb,amsthm,bm,calrsfs,stmaryrd}
\usepackage{hyperref}
\usepackage{times}
\usepackage{mathrsfs}

\makeatletter

\@addtoreset{equation}{section}
\makeatother

\def\bea{\begin{eqnarray}}
\def\eea{\end{eqnarray}}
\def\be{\begin{equation}}
\def\ee{\end{equation}}
\def\ba{\begin{array}}
\def\ea{\end{array}}
\def\ed{\end{document}}

\def\yb{\bar{y}}
\def\zb{\bar{z}}

\def\kappab{\bar{\kappa}}

\def\bea{\begin{eqnarray}}
\def\eea{\end{eqnarray}}
\def\ba{\begin{array}}
\def\ea{\end{array}}
\def\eq#1{(\ref{#1})}

\def\a{\alpha}
\def\b{\beta}
\def\d{\delta}
\def\e{\epsilon}

\def\t{\tau}

\def\L{\Lambda}
\def\O{\Omega}
\def\o{\omega}

\def\ad{\dot{\alpha}}
\def\bd{\dot{\beta}}

\def\una{{\underline{\alpha}}}
\def\unb{{\underline{\beta}}}

\def\ad{{\dot{\alpha}}}
\def\bd{{\dot{\beta}}}

\def\Real{\mathbb{R}}
\def\Comp{\mathbb{C}}

\def\l{\lambda}

\def\m{\mu}

\usepackage{color}
\definecolor{rougef}{rgb}{0.56,0,0}
\definecolor{vertf}{rgb}{0,0.5,0}
\definecolor{bleuf}{rgb}{0,0,0.8}
\definecolor{violetf}{rgb}{0.5,0,0.5}

\begin{document}
\renewcommand{\thefootnote}{\fnsymbol{footnote}}
%

\vspace{7mm}

\begin{center}
{\Large\bf An action principle for Vasiliev's four-dimensional higher-spin gravity}
\vspace{1.5cm}

N i c o l a s~~ B o u l a n g e r \footnote{F.R.S.-FNRS Associate Researcher}~~~~ and ~~~~P e r ~~ S u n d e l l\footnote{F.R.S.-FNRS Researcher with an Ulysse Incentive 
Grant for Mobility in Scientific Research}

\vspace{10mm}
{\footnotesize \tt nicolas.boulanger@umons.ac.be, per.sundell@umons.ac.be}

\vspace*{.5cm}

\textit{Service de M\'ecanique et Gravitation\\Universit\'e de Mons --- UMONS \\20 Place du Parc\\ B-7000 Mons, Belgium}
\vspace*{.5cm}
\end{center}
\vspace{.5cm}
\begin{minipage}{.95\textwidth}
\textsc{Abstract.}
We provide Vasiliev's fully nonlinear equations of motion for bosonic higher spin gauge fields in four spacetime dimensions with an action principle.
We first extend Vasiliev's original system with differential forms in 
degrees higher than one. We then derive the resulting duality-extended 
equations of motion from a variational principle based on a generalized 
Hamiltonian sigma-model action. 
The generalized Hamiltonian contains two types of interaction 
freedoms:
One set of functions that appears in the Q-structure of the generalized 
curvatures of the odd forms in the duality-extended system;  
and another set depending on the Lagrange multipliers, 
encoding a generalized Poisson structure, 
\emph{i.e.} a set of polyvector fields of ranks two or higher in target space. 
We find that at least one of the two sets of interaction-freedom functions must 
be linear in order to ensure gauge invariance.
We discuss consistent truncations to the minimal Type A and B models (with 
only even spins), spectral flows on-shell and provide boundary 
conditions on fields and gauge parameters that are compatible 
with the variational principle and that make the duality-extended system 
equivalent, on shell, to Vasiliev's original system. 
\end{minipage}

\vspace{.9cm}

\renewcommand{\thefootnote}{\arabic{footnote}}

\setcounter{footnote}{0}

\newpage

{\small \tableofcontents }
\newpage

\section{Introduction} \label{sec:Intro}

The natural setting for gauge theories with local space-time symmetries is 
\emph{unfolded dynamics}
\cite{Vasiliev:1988xc,Vasiliev:1990en,Vasiliev:1988sa,Vasiliev:1992gr,Vasiliev:2005zu}. 
The application of this formalism, which is based on \emph{exterior differential systems} (see e.g. \cite{MR1083148,MR2173588} 
and refs. therein), to field theories with local propagating degrees of 
freedom, such as gravities, supergravities and higher-spin gravities, yields infinite towers of zero-forms that are independent dynamical fields off shell. On shell, their integration constants, or expectation values, represent all the local information of the on-shell curvatures, usually referred to as the Weyl tensors. 

In mathematics, an exterior differential system is usually considered as an ideal $I$ in the graded ring of locally defined differential forms on a smooth manifold $M$ that is closed under the operation of exterior differentiation. 
An integral manifold of a differential system is an immersed submanifold of $M$ 
on which each form in $I$ restricts to zero. 
In unfolded dynamics, the generators of $I$ are identified as
\emph{generalized curvatures} and the integral manifold becomes a 
classical solution. Due to Cartan integrability, the curvatures can be integrated and expressed in terms of potentials, providing the fundamental variables in the off-shell formulation.

The canonical framework for the off-shell formulation of unfolded dynamics is based on generalized Poisson sigma models 
\cite{Alexandrov:1995kv,Cattaneo:1999fm,Grigoriev:1999qz,Park:2000au,Ikeda:2000yq,Ikeda:2001fq,Cattaneo:2001ys,Cattaneo:2001bp,Roytenberg:2002nu,Hofman:2002rv,Hofman:2002jz,Ikeda:2002wh},
and 
\cite{Barnich:2005ru,Kazinski:2005eb,Ikeda:2006wd,Grigoriev:2006tt,Roytenberg:2006qz,Lyakhovich:2006sc}. 
Adapting these models to quasi-topological unfolded systems with infinite towers of zero-forms, provides a framework for quantum field theory that one may refer to as unfolded quantum field theory, or deformation quantum 
field theory. They resulting key physical question is whether this novel framework actually contains standard relativistic quantum fields; 
see also \cite{Barnich:2010sw,Grigoriev:2010ic,Kaparulin:2010yr}
for recent developments. \footnote{Note that a relation between the AKSZ 
formalism and unfolding was not explicitly spelled out before 
\cite{Barnich:2005ru}. The observation in \cite{Barnich:2005ru} mainly 
relies on the results of \cite{Barnich:2004cr} where the relation between 
unfolded and BRST approaches was first established (for linear systems).}

Considering retrospectively the works 
\cite{vanNieuwenhuizen:1982zf,D'Auria:1982nx,D'Auria:1982my,D'Auria:1982pm,Castellani:1991eu}, 
one sees that these formulations of supergravities are
examples of unfolded systems, \emph{i.e.} exterior differential systems with 
infinite towers of Weyl zero-forms, though the locality of supergravity implies 
that all the dynamic content can be accessed (in the metric phase) by only 
considering the constraints on the forms in strictly positive degrees, thereby 
explaining why the authors of 
\cite{D'Auria:1982pm,D'Auria:1982my,D'Auria:1982nx} did not consider 
the constraints on the generalized one-form curvatures for the Weyl tensors. 

In this paper we shall address this issue by using the fully non-linear and 
background-independent Vasiliev equations in four spacetime 
dimensions~\cite{Vasiliev:1990en,Vasiliev:1990vu,Vasiliev:1992av}. 
These equations possess an algebraic structure that enables us to construct 
a generalized Hamiltonian action with nontrivial $QP$-structures, and have 
geometric structures which allows to construct additional 
boundary deformations. In this paper we focus on the bulk part 
of the Hamiltonian action, leaving various deformations on 
submanifolds to future works. 
In fact, already in \cite{Vasiliev:1988sa}, such an action principle 
was proposed, which however did not contain any $P$-structure.

We wish to stress that, unlike the original Fronsdal programme, 
which attempts to formulate higher-spin gauge theory off shell 
in a perturbative expansion around constantly curved spacetime, 
the work in this paper provides a background-independent formulation in terms of 
master fields living in the correspondence space, \emph{i.e.} the local product 
of a non-commutative phase-spacetime containing the commutative spacetime as a 
Lagrangian submanifold and a non-commutative twistor space. 
Strictly speaking, the Vasiliev system has a huge classical solution space that 
admits many different perturbative expansions of which only some reduce to 
Fronsdal systems (with cosmological constant).

%

\section{Duality extension on shell}

\subsection{Duality extended bosonic models}

Our starting point is Vasiliev's on-shell formulation of higher-spin gravity 
in four spacetime dimensions 
\cite{Vasiliev:1990en,Vasiliev:1990vu,Vasiliev:1992av}
based on 
combining free differential algebra and the twistor map (see Appendix 
\ref{app:redef}).

Vasiliev's equations of motion provide a particular example of 
formulation of a classical field theory using free differential algebras, 
sometimes referred to as unfolded dynamics. In general, unfolded 
systems can be extended by adding forms in higher degrees. In particular, if the 
underlying differential algebra contains central and closed elements in degrees 
$\{0,2,4,\dots\}\,$, also the structure constants can be extended from the real 
numbers (in degree zero) to general central elements. If this extension is 
nontrivial, that is, if it cannot be removed by a field redefinition, then we 
refer to the resulting extended system as a duality extension of the original 
system. The duality-extended system contains the original system as a consistent 
subsystem, and this subsystem sources the duality-extended sector via nontrivial 
couplings involving central elements of positive degrees (see Appendix 
\ref{app:extension} for a more detailed discussion). 

Vasiliev's equations can be extended adding forms in higher degrees as follows:
\begin{equation}
  A~=~\sum_{p=1,3,\dots}  A_{[p]}\ ,\qquad 
  B~=~\sum_{p=0,2,\dots} B_{[p]}\ ,
\end{equation}
where $ A_{[p]}$ and $ B_{[p]}$ are locally-defined differential 
forms of total degree $p$ belonging to the algebra of bosonic forms 
with generic elements
\begin{equation}
  f~=~\sum_{p=0}^\infty  f_{[p]}(X^M,dX^M;Z^\una, dZ^\una;Y^\una;k,\bar k) \ ,
\end{equation}
\begin{equation}
  f_{[p]}(\lambda\, d X^M;\lambda\, dZ^\una)~=~\lambda^{p}\; 
 f_{[p]}(dX^M;dZ^\una)  \ ,
\end{equation}
for complex parameters $\lambda$ (we suppress the irrelevant variables whenever 
ambiguities cannot arise), where $X^M$ are commuting coordinates, 
$(Y^{\una},Z^{\una})=(y^\a,\yb^{\ad};z^\a,\zb^{\ad})$ are non-commutative 
twistor-space coordinates and $k$ and $\bar k$ are outer Kleinians obeying
\begin{eqnarray}
& k\star  f~=~\pi( f)\star k\ ,\quad 
  \bar k\star  f~=~\bar \pi( f)\star \bar k\ ,\quad 
   k\star k~=~1~=~ \bar k\star \bar k\ ,
\quad &
\end{eqnarray}
with automorphisms $\pi$ and $\bar \pi$ defined by 
$\pi \,{\rm d} ={\rm d} \,\pi\,$, 
$\;\bar\pi \,{\rm d} = {\rm d} \,\bar\pi\;$ and
\bea
& \pi[ f(z^\a,\zb^{\ad};y^\a,\yb^{\ad}) ] ~=~ 
f(-z^\a,\zb^{\ad};-y^\a,\yb^{\ad})\ ,&
\nonumber \\
&\bar \pi[ f(z^\a,\zb^{\ad};y^\a,\yb^{\ad}) ]~=~ 
f(z^\a,-\zb^{\ad};y^\a,-\yb^{\ad})\ .&
\eea
The bosonic projection and irreducibility conditions  
amount to
\begin{equation}
\pi\bar\pi( f) =  f\ ,\qquad f~=~P_+ \star f \ ,\qquad 
{\rm where}\quad  P_\pm ~=~ \frac 12(1 \pm k\star \bar k)\ ,
\end{equation}
which implies
\begin{equation}
f =~ \left[ 
 f^{(+)}(X,dX;Z,dZ;Y) +  f^{(-)}(X,dX;Z,dZ;Y)\star 
\frac{(k+\bar{k})}{2}\,\right] \,\star P_+ \;.
\end{equation}
The bosonic projection removes all component fields associated with the 
unfolding of spinorial degrees of freedom in spacetime.
Irreducible \emph{minimal} 
bosonic models can be obtained by imposing reality 
conditions and discrete symmetries that remove all odd spins; the hermitian 
conjugation $\dagger$ and the relevant anti-automorphism $\tau$ 
are defined by 
$\;{\rm d}[(\cdot)^\dagger] = {[ {\rm d}(\cdot) ]}^\dagger\,$, 
$\;\;{\rm d}\,\tau = \tau\, {\rm d}\;$, 
\begin{eqnarray}
  [ f(z^\a,\zb^{\ad};y^\a,\yb^{\ad};k,\bar k)]^\dagger
 &=& {\bar{f}} (\zb^{\ad},z^{\a};\yb^{\ad},y^{\a};{\bar k},{k})\ ,
 \\[5pt]
 \tau [ f(z^\a,\zb^{\ad};y^\a,\yb^{\ad};k,\bar k)] &=&
 f (-i z^\a,-i \zb^{\ad};i y^\a,i\yb^{\ad};k,\bar k)\ ,
 \\ [5pt]
 {[f_{[p]} \star f'_{[p']}]}^{\dagger} 
  ~=~  {(-1)^{pp'}}{( f'_{[p']})}^{\dagger} \star {( f_{[p]})}^{\dagger}\ 
, &  &
\tau [ f_{[p]}\star f'_{[p']} ] ~=~
{(-1)^{pp'}}\tau( f'_{[p']})\star \tau( f_{[p]})\ . \qquad
\end{eqnarray}
We shall discuss the minimal models below.
\vspace*{.2cm}

The duality extension of the Vasiliev system is based on the following 
generalized curvature constraints
\begin{equation}
  F + {\cal F}~=~0\ ,\qquad 
  D B ~=~0\ ,
\label{dualextend1}
\end{equation}
with Yang--Mills-like curvature and covariant derivative defined by
\begin{equation}
  F~=~ {\rm d} A+ A\star A\ ,\qquad 
  D B~=~ {\rm d}  B+ A\star
  B- B\star A\ ,
\end{equation}
and interaction freedom ($I,\bar I=1,2 $) 
\begin{equation}
 {\cal F}~=~ {\cal F}_I( B)\star  J^I_{[2]} + 
              {\cal F}_{\bar I}( B)\star J^{\bar{I}}_{[2]} 
       +  {\cal F}_{I\bar I} ( B)\star  J_{[4]}^{I\bar{I}}\ 
\end{equation}
featuring the central elements 
\begin{eqnarray}
& ( J^{I}_{[2]})_{I=1,2}~=~-\frac i4(1 \,,\, k \kappa )\,\star P_+\star d^2z  
\;, \quad 
( J^{\bar I}_{[2]})_{\bar I=\bar 1,\bar 2} ~=~ 
-\frac i4( 1\,,\, \bar k {\bar\kappa} )\, \star P_+ \star d^2\bar z  \;,&
\label{central1}\\
&  J^{I\bar I}_{[4]} ~=~  4i\, J^{I}_{[2]} J^{\bar I}_{[2]}\ ,
\label{central2}& 
\end{eqnarray}
and $\star$-functions ${\cal F}_I$, ${\cal F}_{\bar I}$ and 
${\cal F}_{I\bar{I}}$ of $B$ such that  
${\cal F}_I(\l)\,$, ${\cal F}_{\bar I}(\l)\,$ and ${\cal F}_{I\bar I}(\l)\,$ 
($I,\bar I=1,2 \,$), viewed as functions of a single complex variable 
$\l\in \Comp\,$, are complex analytic in a finite neighborhood of $\lambda=0\,$. 
\vspace*{.2cm}

The unfolded equations (\ref{dualextend1}) are Cartan integrable because 
the Yang--Mills-like Bianchi identities ${D} F\equiv 0$ and  
$ D D  B \equiv [ F, B]_\star$ 
are compatible with the generalized curvature constraints.  
In other words, defining the generalized curvatures
\begin{equation}
{\cal R}^A~=~ F+{\cal F} \ ,\qquad {\cal R}^B~=~DB\ ,
\end{equation}
one has the generalized Bianchi identities
\begin{equation}
 D{\cal R}^A-({\cal R}^B\partial_B)\star {\cal F}~\equiv~0\ ,\qquad 
D{\cal R}^B-[{\cal R}^A,B]_\star~\equiv~0\ . 
\end{equation}
The potentials $\{A_{[1]},B_{[2]},A_{[3]},B_{[4]},\dots\}$ in positive form 
degree share one and the same Weyl zero-form $B_{[0]}\,$, that hence contain all 
the local perturbative degrees of freedom of the extended system. One may refer 
to $\{B_{[0]},A_{[1]},B_{[2]},A_{[3]},B_{[4]},\dots\}$ as a duality extension of 
the original Vasiliev system consisting of $\{B_{[0]},A_{[1]}\}$ in the sense 
that the presence of the central elements in degree four implies that 
$\{B_{[2]},A_{[3]},B_{[4]},\dots\}$ cannot in general be set equal to zero on 
shell.  
Moreover, the extension is massless in the sense that for each 
$p\in \{1,2,3,\dots\}$ the system of forms with degrees $p'\leqslant p$ 
constitutes a closed subsystem, \emph{i.e.} their curvatures do not depend on 
the forms with degrees $p'>p\,$. 
In particular, this means that any (locally-defined) exact solution to the 
duality extended system contains a (locally-defined) exact solution to the 
original Vasiliev system. The converse statement requires a more careful 
analysis that we defer here. 

\subsection{A duality extended spectral flow}

The duality extended system possesses a spectral flow \cite{Prokushkin:1998bq} 
describing the evolution of the system on shell under changes in a vacuum 
expectation value $\nu$ and a coupling $g$ defined by the field redefinition
\begin{equation}
  B~=~ \nu{\bf 1}+g B'\ .
\end{equation}
We stress that the parameters $(g,\nu)$ are part of the moduli space of the 
unfolded equations of motion, that is, both $A$ and $B$ depend on $(g,\nu)$ on 
shell and in such a way that the differential ${\rm  d}$ commutes with 
$(\partial_g,\partial_\nu)\,$. 
Letting $f=f(A,dA,B,dB)$ and defining the flow operator 
\begin{equation}
 L_1  f~=~ \partial_g  f-\mu_1  B'\star \partial_\nu  f-\partial_\nu  
f\star \mu_2  B'\ ,\qquad \mu_1,\mu_1~\in~ \Comp\ ,\qquad \mu_1+\mu_2~=~1\ ,
\label{flowL} 
\end{equation}
one has
\begin{eqnarray}
 L_1  F &\equiv&  D L_1 A +\mu_1\,  D B' 
   \star \partial_\nu  A-\mu_2\, \partial_\nu  A \star 
    D B'\ ,\\[5pt] 
  L_1  D  B
   &\equiv&  D L_1 B +[L_1 A, B]_\star + 
   \mu_1\,  D B \star \partial_\nu  B'+\mu_2\, 
   \partial_\nu  B' \star  D B\ ,\\[5pt]
  L_1{\cal F}&\equiv& (L_1 B\partial_{ B})\star {\cal F}\ .
\end{eqnarray}
It follows that the duality extended 
equations of motion are compatible with the flow equations
\begin{equation}
 L_1  A~\approx ~0\ ,\qquad  L_1  B~\approx~0\ ,\label{flow} 
\end{equation}
where the last flow equation is equivalent to that $L_1 B'\approx0\,$. 
\vspace*{.2cm}

The flow equations generalize as follows: one first redefines
\be  B = \nu + {\cal N}( B')\ ,\qquad {\cal N} = \nu_1 g  B' + \nu_2 g^2  
B'^{\star2} + \nu_3 g^3  B'^{\star 3} + \cdots\ ,
\ee
where $\nu_k$ ($k\geqslant 1$) are constants and $g$ the coupling. 
The flow operator defined by
\be L  f~=~ \partial_g  f - {\cal M}_1( B') \star\partial_\nu  f 
         - \partial_\nu  f \star {\cal M}_2( B')\ ,
\ee
where the two $\star$-functions defined by ($i=1,2$) 
\be {\cal M}_i~=~\mu_{i,1}\, g \,  B' + \m_{i,2} \, g^2 \, B'^{\star 2} 
   + \ldots \ ,
\qquad      \m_{1,k} + \m_{2,k} ~=~ k \, \nu_k\quad (k\geqslant 1)\ ;
\ee
obey
\bea L {\cal F} &\equiv& (L B\partial_{ B})\star {\cal F}\ ,\\[5pt]
L B& =& \nu_1 L B' + \nu_2 g^2 (L B'\star  B' +  B'\star L B') + \cdots\ ,
\\[5pt]
L F& =&  D L A +  D{\cal M}_1\star \partial_\nu  A - \partial_\nu  
A\star  D{\cal M}_2\ ,
\\[5pt]
L D B'& =&  DL B' + [L A, B']_\star  +  D{\cal M}_1\star \partial_\nu  B' + 
\partial_\nu B'\star D{\cal M}_2\ ,
\eea
and it follows that one can set the constraints
\be  
L A ~=~ 0\ ,\qquad L B'~=~ 0\ ,
\ee
where the latter constraint thus implies that $L B=0\,$.
One can redefine ${\cal N}=g  B'$ so that $\nu_1=1$ and $\nu_k=0$ for $k>1$, 
leaving the freedom in ${\cal M}_i$ that generalizes the two-parameter freedom 
in  having $\mu_1$ and $\mu_2\,$.  

\subsection{Consistent truncations}\label{sec:truncations}

There are two possible reality conditions leading to models with
negative cosmological constant $\L<0\,$, 
that we parameterize using $\e_{\Real}=\pm1$ as follows: 
\begin{equation} 
(A_{[p]})^\dagger~=~-(\e_{\Real})^{\frac{p-1}2}A_{[p]}\ ,\qquad 
   ( B_{[p]})^\dagger~=~ (\e_{\Real})^{\frac{p}2} B_{[p]}\ ,\ee
\be ({\cal F}_I(\l))^\dagger~=~ {\cal F}_{\bar I}(\l^\dagger)\ ,\qquad 
{\cal F}_{I\bar J}(\l))^\dagger~=~\e_{\Real} \,{\cal F}_{J\bar I}(\l^\dagger)\ .
\end{equation}
Moreover, using the map 
\begin{equation}
\pi_k \; : \; (k,\bar k)\;\mapsto\; (-k,-\bar k)\ ,  
\end{equation}
there are two possible projections to models without topological (adjoint) 
zero-forms, that we parameterize using $\epsilon_k = \pm 1$ as follows:
\begin{equation}
 \pi_k( A_{[p]})~=~(\epsilon_k)^{\frac{p-1}2} A_{[p]}\ ,\qquad 
 \pi_k( B_{[p]})~=~-(\epsilon_k)^{\frac{p}2} B_{[p]}\ ,
\end{equation}
\begin{equation}
 {\cal F}_I(-\l)~=~(-1)^{I+1} {\cal F}_I(\l)\ ,\qquad 
{\cal F}_{I\bar I}
  (-\l)~=~(-1)^{I+\bar I}\e_k \, {\cal F}_{I\bar I}(\l)\ .
\end{equation}
Using the parity transformation $P$ defined by 
$P \,{\rm d}= {\rm d} \, P$ and 
\begin{equation}
 P\left[ f(X^M;z^\a,\zb^{\ad};y^\a,\yb^{\ad};k,\bar k) \right]~=~
(P f)(X^M;-\zb^{\ad},-z^\a;\yb^{\ad},y^\a;\bar k,k)\ ,
\end{equation}
which is an automorphism of the $\star$-product algebra and where $P f$ is 
expanded in terms of parity reversed component fields, there are four ways of 
fixing parities, that we parameterize using $\e,\tilde \e=\pm 1$ as follows:
\be 
P( A_{[p]})~=~(\e\tilde\e)^{\frac{p-1}2} A_{[p]}\ ,\qquad 
P( B_{[p]})~=~(\e)^{\frac{p+2}2}(\tilde\e)^{\frac{p}2} B_{[p]}\ ,
\ee
\be {\cal F}_{\bar I}(\l)~=~{\cal F}_I(\e \l)\ ,\qquad {\cal F}_{I\bar J}
(\l)~=~\e\tilde\e {\cal F}_{J\bar I}(\e\l)\ .
\ee
Finally, the $\tau$-projection to the minimal models with only even propagating 
spins reads
\begin{equation}
 \tau( A_{[p]})~=~(-1)^{\frac{p+1}2}  A_{[p]}\ ,\qquad 
                    \tau( B_{[p]})~=~(-1)^{\frac{p}2}  B_{[p]}\ , 
\end{equation}
which is the unique choice since $\tau( J_{[p]})=(-1)^{\frac{p}2} J_{[p]}$ 
(and there is no condition on ${\cal F}$). 

\vspace*{0.3cm}
In the $( B_{[0]}, A_{[1]})$-sector, which forms a closed 
subsystem, the assignement of $k$-parity combined with the freedom in redefining 
$ A_\a$ can be used to replace 
\cite{Vasiliev:1990en} 
%
\begin{equation}
 ({\cal F}_1,{\cal F}_2;{\cal F}_{\bar 1},{\cal F}_{\bar 2})~\rightarrow~ 
 (0,(1-{\cal F}_1)^{\star(-1)}\star {\cal F}_2;0,
      (1-{\cal F}_{\bar 1})^{\star(-1)}\star {\cal F}_{\bar 2})\ .
\end{equation}
Imposing also reality and parity conditions, of which the latter is a multiple 
choice parametrized by $\epsilon=\pm 1\,$, the remaining interaction function 
$(1-{\cal F}_1)^{\star(-1)}\star {\cal F}_2$ becomes real and odd, hence 
defining the new master field 
\begin{equation}
\Phi\star P_+ ~=~ (1-{\cal F}_1)^{\star(-1)}\star {\cal F}_2 \star 
k \star P_+\ , 
\end{equation}
obeying the twisted reality condition $(\Phi)^\dagger=\pi( \Phi)\,$ and the 
parity condition $P(\Phi)=\e\, \Phi$ leading to a physical scalar that is even 
under parity for $\e=1$ and odd under parity for $\e=-1$. Finally, one may 
project out the odd spins by imposing $\tau( \Phi)=\pi( \Phi)$ yielding the 
minimal bosonic models.

Assuming linear interaction functions 
\begin{equation}
 {\cal F}_I = b_I \, B\quad, \qquad
 {\cal F}_{\bar I} = b_{\bar I} \, B\quad, \qquad
 {\cal F}_{I\bar I} = c_{I\bar I} \, B\quad, 
\end{equation}
and defining a total central element 
\begin{equation}
J~=~ J_{[2]}+ J_{[4]}
\label{defJ} 
\end{equation}
via 
\begin{equation}
  B \star  J_{[2]}~=~ {\cal F}_I \star J^I_{[2]}
  + {\cal F}_{\bar I}\star  J^{\bar I}_{[2]}  \quad , \qquad 
   B \star J_{[4]} 
   ~=~  {\cal F}_{I\bar I}\star J^{I\bar I}_{[4]}\ ,  
\end{equation}
\begin{equation}
 J_{[2]}~=~ -\frac{i}4 \, \left[ {\rm d}z^2 ( b_1+b_2 \,k\,\kappa) +
                  {\rm d}\bar z^2 (b_{\bar 1}+ b_{\bar 2}\, \bar k \, 
                   {\bar\kappa})\ \right]\star \, P_+\;,
\label{J2} 
\end{equation}
\begin{equation}
 J_{[4]}~=~ -\frac{i}4\, {\rm d}z^2 {\rm d}\zb^2 
         \left[ c_{1\bar 1} + 
            c_{2 \bar 1 }\, k\,\kappa + c_{1 \bar 2}\, \bar k\,
                 {\bar \kappa} + c_{2\bar 2}\,  \kappa 
              \,{\bar \kappa}\right] \star \, P_+ \quad,
\label{J4} 
\end{equation}
the reality, $k$-parity and $P$-parity conditions imply
\be ( J_{[p]})^\dagger~=~-(\e_{\Real})^{\frac{p-2}2} J_{[p]}\ ,\qquad 
\pi_k( J_{[p]}) ~=~-(\epsilon_k)^{\frac{p-2}2} J_{[p]}\ ,\qquad P( J_{[p]})
~=~
(\e)^{\frac{p}2}(\tilde\e)^{\frac{p-2}2} J_{[p]}\ ,
\ee
which constrain the parameters $(b_I,b_{\bar I}, c_{I\bar I})\,$.
These conditions admit nontrivial solutions for $J_{[p]}$ for all combinations 
of signs except for $\e_k=\tilde \e=-1$ since $\e_k=-1$ implies that 
$\tilde \e=+1\,$. 

\section{Generalized Hamiltonian action principle}

\subsection{Graded cyclic chiral trace}

Vasiliev's equations are formulated in terms of master fields which one may
think of as functions on a total space called \emph{correspondance space} 
$\mathfrak{C}\,$, that is locally a product space 
$M_{\xi}\times {\mathfrak{Z}}\times {\mathfrak{Y}}$ 
where ${\mathfrak{Z}}$ and $\mathfrak{Y}$ are two copies of a 
non-commutative twistor space and
$M_\xi$ denotes a coordinate chart of a commuting base manifold $M\,$, 
see Appendix \ref{app:redef} for more details.  
In order to build an action principle, we need to integrate over the
correspondance space. 
The integration over ${\mathfrak{C}}$ of a globally defined $(\hat p+1)$-form 
${\cal L}$ is defined by
\begin{equation}
 \int_{{\mathfrak{C}}} {\cal L}~=~\sum_{\xi} \int_{M_\xi}  {\rm Tr} 
\left[ f_{\cal L}\right]\ ,
\end{equation}
where $ f_{\cal L}$ denotes a symbol of ${\cal L}$ and the chiral trace 
operation is defined by
\begin{equation}
{\rm Tr} \left[ f\right] ~=~\sum_m \int_{{\mathfrak{Z}}\times {\mathfrak{Y}}}  
 \frac{d^2y \,d^2\yb} {(2\pi)^2}\;
 \frac{f_{[m;2,2]}|_{k= 0 = \bar k}}{(2\pi)^2}\ , 
\end{equation}
using the decomposition 
$ f_{[p]}=\sum_{\tiny \ba{c} m+q+\bar q=p\\ q,\bar q\leqslant 2\ea}  f_{[m;q,\bar q]}\;$ 
with 
\begin{equation}
f_{[m;q,\bar q]}(\lambda \,{\rm d}X^M; \mu \, {\rm d}z^\alpha,
\bar{\mu} \, {\rm d}\zb^{\ad})~=~
\lambda^m \,\mu^q \,\bar\mu^{\bar q}\, f_{[m;q,\bar q]} ( {\rm d}X^M ; {\rm d}z^\a, 
{\rm d}\zb^{\ad})\ , 
\end{equation}
and with integration domain consisting of real contours for $\{y^\a,z^\a\}$ and 
$\{ \bar y^\ad , \bar z^\ad \}\,$, respectively, that is, one performs separate 
integrations over the holomorphic and anti-holomorphic variables treated as 
independent real variables (for related discussions, see \emph{e.g.} Appendix G 
of \cite{Iazeolla:2008ix}).
The choice of the chiral integration domain (instead of the complex integration 
domain) implies that  
\begin{equation}
  {\rm Tr}\left[\pi( f)\right]~=~ {\rm Tr}\left[\bar\pi( f)\right]\ =\ 
  {\rm Tr}\left[ f\right]\ ,\label{chiralsplit}
\end{equation}
which in its turn implies graded cyclicity,
\begin{equation} 
 {\rm Tr}\left[ f_{[p]}\star  f'_{[p']}\right]~=~ (-1)^{pp'}\;
 {\rm Tr}\left[ f'_{[p']}\star  f_{[p]}\right]\ ,
\label{gradedcyclic}
\end{equation}
as can seen by expanding 
$ f_{[p]}=( f^{(+)}_{[p]}+f^{(-)}_{[p]}\star k)\star P_+\;$ \emph{idem} 
$\;f'_{[p']}\,$ which yields 
\begin{equation}
  {\rm Tr}\left[ f_{[p]}\star  f'_{[p']}\right]~=~\frac12  \,{\rm Tr}
\left[ f^{(+)}_{[p]}\star  f^{\prime(+)}_{[p']}+ f^{(-)}_{[p]}\star 
\pi( f^{\prime(-)}_{[p']})\right]\ ,
\end{equation}
where the second term is graded cyclic by virtue of the chiral integration. 
Furthermore, the chiral trace operation commutes to hermitian conjugation and is 
invariant under $P$ and $\pi_k\,$,
\be 
\left( {\rm Tr}\left[ f\right]\right)^\dagger~=~{\rm Tr}
\left[( f)^\dagger\right]\ ,\qquad  {\rm Tr}\left[P( f)\right]~=~
{\rm Tr}\left[ f\right]\ ,\qquad  {\rm Tr}\left[\pi_k( f)\right]~=~
{\rm Tr}\left[ f\right]\ .
\ee
Finally, one may seek to impose boundary conditions in 
${\mathfrak{Z}} \times {\mathfrak{Y}}$ such that the integration contours can 
be rotated from real to imaginary axes in the sense that 
\be  
{\rm Tr}\left[\tau( f)\right]~=~{\rm Tr}\left[ f\right]\ .
\ee
We shall finally assume that the integration over ${\mathfrak{C}}$ 
is non-degenerate 
such that if ${\rm Tr}\left[ f\star g\right]=0$ for all $f$ then $g=0\,$.  
It is an interesting open problem to understand  whether the $\pi$, $P$ and 
$\tau$ symmetries could be violated on classical observables evaluated on 
exact solutions that one may seek to 
interpret as describing topology changes of the twistor space which we leave for 
future studies \cite{wipCarlo}. 
In what follows, we shall always assume that the discrete 
symmetries hold off shell. 
 
\subsection{Odd-dimensional bulk ($\hat p\in 2\mathbb N$)}

\subsubsection{Action principle}

In the case of an odd-dimensional base manifold of dimension  
$\hat p+1= 2n+5$ with $n\in\{0,1,2,\dots\}$ such that 
${\rm dim}(M)=2n+1\,$, the duality-extended equations of motion follow 
from the variational principle based on the generalized Hamiltonian bulk action
\begin{eqnarray}
S^{\rm cl}_{\rm bulk}[\{ A, B, U, V\}_\xi]
&=& \sum_\xi \int_{M_\xi}  {\rm Tr}\left[
 U \star D B+ V\star \left( F + {\cal G}(B,U;  J^I,  J^{\bar I}, 
J^{I\bar I})\right)\right]\ ,\qquad
\label{SclHS} 
\end{eqnarray}
with interaction freedom ${\cal G}$ and locally-defined master fields 
decomposing under total form degree into 
\begin{equation}
    A  ~=~   A_{[1]}+   A_{[3]}+  \cdots +   A_{[2m-1]}\ ,\qquad 
    B  ~=~   B_{[0]}+   B_{[2]}+ \cdots +   B_{[2m-2]}\ ,\ee\be
    U  ~=~   U_{[2]}+   U_{[4]}+ \cdots +   U_{[2m]}\ ,\qquad  
    V  ~=~    V_{[1]}+   V_{[3]}+ \cdots +   V_{[2m-1]}\ , \quad m=n+2\ .
\end{equation}
The function ${\cal G}$ must be constrained in order for the action to be gauge 
invariant and in order to avoid systems that are trivial. In what follows, we 
shall consider the special case 
\be 
{\cal G}~=~{\cal F}(B;  J^I,  J^{\bar I}, J^{I\bar I})+
\widetilde{\cal F}(U;  J^I,  J^{\bar I}, J^{I\bar I})\ ,
\ee
\bea
 {\cal F}&=& {\cal F}_I( B)\star  J^I_{[2]} + 
              {\cal F}_{\bar I}( B)\star J^{\bar{I}}_{[2]} 
       +  {\cal F}_{I\bar I} ( B)\star  J_{[4]}^{I\bar{I}}\ ,\\[5pt]
\widetilde{\cal F}&=& \widetilde{\cal F}_0(U)+\widetilde{\cal F}_I( U)
\star  J^I_{[2]} + 
              \widetilde{\cal F}_{\bar I}( U)\star J^{\bar{I}}_{[2]} 
       +  \widetilde{\cal F}_{I\bar I} ( U)\star  J_{[4]}^{I\bar{I}}\ ,
\eea
where the (non-)vanishing of the coupling
$\lambda:=\partial_U \widetilde{\cal F}_0\vert_{U=0}\,$ implies that
the target space is equipped with a Poisson (symplectic) structure. 
In the case of a proper Poisson structure with $\lambda=0\,$ the action cannot
be written as a boundary term.

Denoting $Z^{ i}=( A, B, U, V)$\,, the general variation of the action defines 
generalized curvatures ${\cal R}^i$ as follows:
\begin{equation}
\delta S\ =\ \sum_\xi \int_{M_\xi}  {\rm Tr}\left[ {\cal R}^{ i} \star 
\delta  Z^{ j} {\cal O}_{ i j}
\right]+ \sum_{\xi} \int_{\partial M_\xi}  {\rm Tr} \left[  U\star\delta  B
      - V\star\delta  A \right]\ , \label{totalvar}
\end{equation}
where one thus has
\be
{\cal R}^{ A} ~=~   F + {\cal F}+ \widetilde{\cal F}\ ,\qquad 
{\cal R}^{ B} ~=~   D  B + ( V\partial_{ U})\star \widetilde{\cal F}\ ,\ee
\be {\cal R}^{ U} ~=~   D  U -( V\partial_{ B})\star {\cal F}\ ,\qquad 
{\cal R}^{ V} ~=~  D  V + [ B ,  U]_\star \ ,\ee
with ${\cal O}_{ i j}$ being a constant non-degenerate matrix (defining a 
symplectic form of degree $\hat p+2$ on the $\mathbb N$-graded target space of 
the bulk theory). 
Treating $Z^i$ and $dZ^i$ as independent variables, one has the differential 
identities
\bea  D {\cal R}^{ A} - ({\cal R}^{ B}\partial_{ B})\star {\cal F}
-({\cal R}^{ U}\partial_{ U})\star \widetilde{\cal F}&\equiv& {\cal A}^{ A}\ ,
\\[5pt]
 D {\cal R}^{ B} - [{\cal R}^{ A}, B]_\star-({\cal R}^{ V}\partial_{ U})
\star \widetilde{\cal F} -  ({\cal R}^{ U}\partial_{ U})\star 
( V\partial_{ U})\star \widetilde{\cal F}&\equiv& {\cal A}^{ B}\ ,
\\[5pt]
 D {\cal R}^{ U} - [{\cal R}^{ A}, U]_\star+({\cal R}^{ V}\partial_{ B})
\star {\cal F}+  ({\cal R}^{ B} \partial_{ B})\star ( V\partial_{ B})
\star \widetilde{\cal F}&\equiv& {\cal A}^{ U}\ ,
\\[5pt]
 D {\cal R}^{ V} - [{\cal R}^{ A}, V]_\star
- [{\cal R}^{ B}, U]_\star + [{\cal R}^{ U}, B ]_\star  &\equiv&{\cal A}^{ V}\ ,
\eea
with $dZ^i$-independent quantities ${\cal A}^i\equiv {\cal A}^i(Z^j)$ given by
\bea {\cal A}^{ A}&\equiv&-(( V\partial_{ U})\star \widetilde{\cal F})
\partial_{ B}\star {\cal F}+(( V\partial_{ B})\star {\cal F})\partial_{ U}
\star \widetilde{\cal F}\ ,
\\[5pt] 
{\cal A}^{ B}&\equiv&(( V\partial_{ B})\star {\cal F})\partial_{ U}\star 
( V\partial_{ U})\star\widetilde{\cal F}\ ,
\\[5pt]
{\cal A}^{ U}&\equiv&(( V\partial_{ U})\star \widetilde{\cal F})
\partial_{ B}\star ( V\partial_{ B})\star{\cal F}\ ,
\\[5pt]
{\cal A}^{ V}&\equiv&0\ ,
\eea
where the last identity follows from
\be [ U,( V\partial_{ U})\star \widetilde{\cal F}]_\star~\equiv~-
[ V,\widetilde{\cal F}]_\star\ ,\qquad [ B,( V\partial_{ B})\star 
{\cal F}]_\star~\equiv~-[ V,{\cal F}]_\star\ .
\ee
The quantities ${\cal A}^i$ thus represent obstructions to generalized Bianchi 
identities off shell and hence to Cartan integrability of the unfolded equations 
of motion ${\cal R}^i\approx 0\,$, where in this Section we use weak equalities
for equations that hold on shell. 
These obstructions vanish identically (without further algebraic constraints 
on $Z^i$) in at least the following two cases:
\bea \mbox{bilinear $Q$-structure}&:&{\cal F}~=~  B\star  J\ ,\qquad 
J~=~J_{[2]}+ J_{[4]}\ ,\\[5pt]
\mbox{bilinear $P$-structure}&:& \widetilde{\cal F}~=~  U\star  J'\ ,\qquad  
J'~=~J'_{[2]}+ J'_{[4]}\ ,
\eea
where the central elements are expanded as in Eqs. \eq{defJ}--\eq{J4}. 

At this stage it is useful to recall (see Appendix \ref{app:c}) that if 
${\cal R}^i=dZ^i+{\cal Q}^i(Z^j)$ defines a set of generalized curvatures, 
then one has the following three equivalent statements: (i) ${\cal R}^i$ obey a 
set of generalized Bianchi identities 
$d{\cal R}^i-({\cal R}^j\partial_j)\star {\cal Q}^i\equiv 0$; (ii) ${\cal R}^i$ 
transform into each other under Cartan gauge transformations 
$\delta_\varepsilon Z^i=d\varepsilon^i-(\varepsilon^j\partial_j)\star {\cal Q}^i\,$; 
and (iii) the quantity $\overrightarrow{\cal Q}:={\cal Q}^i\partial_i$ is a 
$Q$-structure, \emph{i.e.} a nilpotent $\star$-vector field of degree one in 
target space, \emph{viz.} $\overrightarrow{\cal Q}\star {\cal Q}^i\equiv 0\,$. 
Furthermore, in the case of differential 
algebras on commutative base manifolds, one can show that if ${\cal R}^i$ are 
defined via a variational principle as in \eq{totalvar} (with constant 
${\cal O}_{ij}$), then the action $S$ remains invariant under 
$\delta_\varepsilon Z^i\,$. 

In the two Cartan-integrable cases at hand, one thus has the on-shell Cartan gauge 
transformations
\bea \delta_{ \e,\eta}  A&=& D \e^{\, A}-(\e^{\, B}\partial_{ B})\star {\cal F}-(\eta^{\, U}\partial_{ U})\star\widetilde{\cal F}\ ,\\[5pt]
\delta_{ \e,\eta}  B&=& D \e^{\, B}-[ \e^{\, A}, B]_\star-( \eta^{\, V}\partial_{ U})\star \widetilde{\cal F}-( \eta^{\, U}\partial_{ U})\star( V\partial_{ U})\star\widetilde{\cal F}\ ,\\[5pt]
 \delta_{ \e,\eta}  U&=& D \eta^{\, U}-[ \e^{\, A}, U]_\star+( \eta^{\, V}\partial_{ B})\star {\cal F}+( \e^{\, B}\partial_{ B})\star( V\partial_{ B})\star{\cal F}\ ,\\[5pt]
\delta_{ \e,\eta}  V&=& D \eta^{\, V}-[ \e^{\, A}, V]_\star-[ \e^{\, B}, U]_\star +[ \eta^{\, U}, B]_\star\ .\eea
These transformations remain symmetries off shell as can be seen using the following set of identities:
\bea \mbox{bilinear $P$-structure}&:& {\rm Tr}\left[ J'\star  V\star  
( V\partial_{ B})\star( \e^{\, B}\partial_{ B})\star{\cal F}\right]~\equiv~0\ ,
\\[5pt]
&&{\rm Tr}\left[ V\star ( D B\partial_{ B})\star ( \e^{\, B}
\partial_{ B})\star{\cal F}+ D B\star ( V\partial_{ B})\star 
( \e^{\, B}\partial_{ B})\star{\cal F}\right]~\equiv~0\ ,
\nonumber\\[5pt]
&&{\rm Tr}\left[\eta^{\, V}\star ( D B\partial_{ B})\star{\cal F}- 
D B\star ( \eta^{\, V}\partial_{ B})\star{\cal F}\right]~\equiv~0\ ,
\\[10pt]
\mbox{bilinear $Q$-structure}&:&{\rm Tr}\left[ J\star  V\star  
( V\partial_{ U})\star( \eta^{\, U}\partial_{ U})\star
\widetilde{\cal F}\right]~\equiv~0\ ,
\\[5pt]
&&{\rm Tr}\left[ V\star ( D U\partial_{ U})\star ( \eta^{\, U}
\partial_{ U})\star\widetilde{\cal F}+ D U\star ( V\partial_{ U})\star 
( \eta^{\, U}\partial_{ U})\star\widetilde{\cal F}\right]~\equiv~0\ ,
\nonumber \\[5pt]
&&{\rm Tr}\left[\eta^{\, V}\star ( D U\partial_{ U})\star\widetilde{\cal F}- D U\star ( \eta^{\, V}\partial_{ U})\star\widetilde{\cal F}\right]~\equiv~0\ .\
\eea
More precisely, the $(\e^{\, A} , \e^{\, B})$-symmetries leave the Lagrangian 
invariant while the $( \eta^{\, U}, \eta^{\, V})$-symmetries transform the 
Lagrangian into a nontrivial total derivative, \emph{viz.}
\be 
\delta_{\e,\eta}{\cal L}~\equiv~ {\rm d}\left(Tr\left[\eta^U \star 
{\cal K}_U+\eta^V\star  {\cal K}_V\right]\right)\ ,\label{epsilonident}
\ee
for $({\cal K}_U,{\cal K}_V)$ that are not identically zero.
It follows that the Cartan gauge algebra $\mathfrak{g}$ is of the form 
$$\mathfrak{g}\cong {\mathfrak g_1}\subsetplus {\mathfrak g}_2$$ 
with ${\mathfrak g_1} \cong {\rm{span}}\{ \e^{\, A}, \e^{\, B} \}$ and 
${\mathfrak g_2} \cong {\rm{span}}\{ \eta^{\, U}, \eta^{\, V} \}\,$, 
as one can verify explicitly using the formulae \eq{closure} 
given in Appendix \ref{app:c}.  


\subsubsection{Global formulation, boundary conditions and embedding of
Vasiliev's original system}\label{relation}

Exponentiation of the infinitesimal Cartan gauge transformations leads to 
locally-defined gauge orbits consisting of elements (see Appendix \ref{app:global})
\be  Z^{ i}_{ \l,{\rm d}\l;Z_0}~=~{\cal G}_{\l,{\rm d}\l;Z}\star Z^i|_{Z^i= Z^{ i}_0}\ ,
\ee
\be {\cal G}_{\l,{\rm d}\l;Z}~:=~\exp_\star \overrightarrow{\cal T}_{\l,{\rm d}\l;Z}\ ,
\qquad  
\overrightarrow{\cal T}_{\l,{\rm d}\l;Z}~:=~\left( {\rm d} \, \l^{ i}-
( \l^{ j}\partial_{ j})\star {\cal Q}^{ i}\right)\frac{\partial}{\partial Z^i}\ ,
\ee
where $ \lambda^{ i}$ and $Z^i_0$, respectively, are gauge functions and representatives of 
the orbits defined in coordinate charts of the base manifold. 
On shell, one has
\be  {\rm d}\, Z^{ i}_0+{\cal Q}^{ i}( Z^{ j}_0)~\approx~0\quad \Rightarrow\quad  
{\rm d}\, Z^{ i}_{ \l,d\l;Z_0}+{\cal Q}^{ i}( Z^{ j}_{ \l,d\l;Z_0})~\approx~0\ ,
\ee
as can be seen by first writing 
${\rm d}\approx\overrightarrow{\cal S}_{ {\rm d} \l}- \overrightarrow{\cal Q}\,$ 
where 
$\overrightarrow{\cal S}_{ d \l}:=  {\rm d}\lambda^{ i} \partial/\partial \l^i\,$ 
and $\overrightarrow{\cal Q}:={\cal Q}^{ i} \partial/\partial Z^i\,$, 
and then using 
$\left[\overrightarrow{\cal S}_{ {\rm d} \l}- \overrightarrow{\cal Q}, 
\overrightarrow{\cal T}_{ \l,{\rm d}\l;Z}\right]_\star \equiv 0\,$ 
and 
$\left[\exp_\star \overrightarrow{\cal X}\right]\star \left({\cal F}\star
{\cal F'}\right)\equiv\left(\left[\exp_\star \overrightarrow{\cal X}\right]\star 
{\cal F}\right)\star \left(\left[\exp_\star \overrightarrow{\cal X}\right]\star 
{\cal F'}\right)\,$ 
for any 
$\star$-vector field $\overrightarrow{\cal X}$  and $\star$-functions ${\cal F}$ and 
${\cal F}'$ (see Appendix \ref{app:c} for details).

In particular, it follows that the space of (locally-defined) classical solutions to the 
duality extended $( A, B; U, V)$-system contains a subspace of (locally-defined) classical 
solutions to the duality extended $( A, B)$-system, obtained simply by setting 
$U=0= V\,$. 
The $( A, B)$-system contains in its turn a subset of the (locally-defined) solutions to the 
original Vasiliev system in form degrees $0$ and $1\,$. 
The converse issue, whether any given (locally-defined) exact solution to the original 
Vasiliev system can be uplifted to the $( A, B)$-system, requires, however, a more careful 
analysis of the gauge orbits in degrees greater than $1$ (due to the non-polynomial 
dependencies on the integration constants for the Weyl zero-form and the zero-form gauge 
functions).

Turning to the global formulation, it follows from Eq. \eq{epsilonident} that the gauge 
parameters $(\e^A_\xi,\e^B_\xi)\in\mathfrak g_1$ can be locally defined on $M\,$, that is, 
defined independently on the coordinate charts $M_\xi\,$ --- 
provided that the action is not perturbed by impurities that break some of the 
$(\e^A,\e^B)$-symmetries, as for example in the soldered phase where perturbations break 
the local translations in $\e^{A_{[1]}}\,$.
From Eq. \eq{epsilonident} it also follows that $(\eta^U,\eta^V)\in\mathfrak g_2$ need to be 
defined globally on $M\,$, that is, $(\eta^U,\eta^V)|_\xi$ and $(\eta^U,\eta^V)|_{\xi'}$ 
must be related by transition functions $\{t^{\xi'}_{\xi}\}\,$
across the chart boundary between $M_\xi\,$ and $M_{\xi'}\,$; in practice this means that one 
may take $(\eta^{\, U}_\xi,\eta^{\, V}_\xi)$ to have compact support in $M_\xi\,$.

The unbroken phase of the theory thus consists of local representatives 
$Z^i_\xi=(A,B;U,V)|_\xi$ defined up to gauge transformations with parameters 
$( \e^{\, A}_\xi;\e^{\, B}_\xi)$ that are unrestricted on $\partial M_\xi$ and parameters 
$(\eta^{\, U}_\xi,\eta^{\, V}_\xi)$ with the aforementioned restrictions on 
$\partial M_\xi\,$, with transitions of the form
\be  Z^{ i}_\xi~=~{\cal G}_\xi^{\xi'}\star   Z^{ i}_{\xi'}\qquad 
\mbox{defined on\quad $M_\xi\cap M_{\xi'}$}\ ,
\label{transitions}
\ee
where ${\cal G}_\xi^{\xi'}=\exp_\star \overrightarrow{\cal T}_{t,{\rm d} t;Z}|_{\xi}^{\xi'}\,$ 
with transition functions $t_\xi^{\xi'}\in \mathfrak g_1$ defined on $M_\xi\cap M_{\xi'}\,$.

More generally, softly broken phases of the theory arise by taking the transition 
functions $ \{t_\xi^{\xi'}\}$ to be generated by various unbroken subalgebras 
$ {\mathfrak l}\subseteq {\mathfrak g}_1\,$. Their moduli spaces ${\cal M}_{ {\mathfrak l}}$ can be coordinatized by classical observables 
${\cal O}_{ {\mathfrak l}}$ that are manifestly $ {\mathfrak l}$-invariant off shell and 
diffeomorphism-invariant on shell (one may thus think of the unbroken phase 
${\cal M}_{ {\mathfrak g}}$ as the smallest homotopy phase for a given base manifold; 
it can be embedded into various broken phases). Of particular interest is the soldered phase 
in which the action is perturbed as to softly break the gauge symmetries associated with the 
$\pi$-odd projection of $A_{[1]}\,$. 
The unbroken gauge algebra in this case thus consists of the $\pi$-even projection 
$\frac12(1+\pi)\e^{A_{[1]}}$ together with the remaining $\e$-parameters of positive form 
degree.

Hence, to achieve a globally well-defined variational principle, one considers 
globally-defined field configurations off shell consisting of locally-defined 
representatives $\{Z^{ i}_\xi\}$ 
related on chart boundaries via transitions (\ref{transitions}) 
for a given structure algebra $ {\mathfrak l}\subseteq {\mathfrak{g}}_1\,$. 
The manifest $\mathfrak{ g}_1$-invariance implies that in the general variation 
(\ref{totalvar}), the contributions from two adjacent boundaries $\partial M_\xi$ and 
$\partial M_{\xi'}$ cancel; on such a boundary one has the transition functions 
($t \equiv t_\xi^{\xi'}$) 
\begin{eqnarray}
 \delta_{ t}(\delta  A)&=&-[t^{\, A},\delta A]_\star- (\delta B\partial_{ B})\star 
(t^{\, B}\partial_{ B})\star{\cal F}\ ,
\\[5pt]
\delta_{ t}(\delta  B)&=&-[t^{\, A},\delta B]_\star+\{t^{\, B}, \delta A\}_\star\ ,
\\[5pt]
\delta_{ t}  U&=&-[t^{\, A}, U]_\star+  (t^{\, B}\partial_{ B})\star 
( V\partial_{ B})\star{\cal F}\ ,
\\[5pt]
\delta_{ t}  V&=&-[t^{\, A}, V]_\star-[t^{\, B}, U]_\star\ , 
\end{eqnarray}
which implies that ($t \equiv t_\xi^{\xi'}$)
\be 
\delta_{t}\left(\int_{\partial M_\xi}{\rm Tr}\left[ 
U\star \delta B- V\star\delta A\right]\right)
\ee
\be
 =~ \int_{\partial M_\xi}{\rm Tr}\left[ V\star(\delta B\partial_{ B})\star 
(t^{\, B}\partial_{ B})\star {\cal F}-\delta B \star( V\partial_{ B})\star 
(t^{\, B}\partial_{ B})\star {\cal F}\right]~\equiv~0\ .
\ee
One is thus left with contributions from true boundaries $\partial M_\xi\subset \partial M$ 
(including boundaries of homotopy cylinders surrounding impurities of co-dimension greater 
than one). It follows that the natural boundary conditions compatible with the 
locally-defined gauge symmetries are the Dirichlet conditions
\be ( U, V)|_{\partial M}~=~ 0 \ .\label{BCUV}
\ee
In summary, a classical solution can thus be specified by fixing 
\begin{itemize}
 \item[(i)] the transition functions 
$\{t_\xi^{\xi'}\} \in {\mathfrak l}\subseteq {\mathfrak g}_1\,$;  
\item[(ii)] an initial datum for the zero-form $B_{[0]}\,$, say
\be  
B_{[0]}|_{p}~=~C(Y;k,\bar k)\ ,
\ee
at some given point $p\in \mathfrak B$ in the base manifold; 
\item[(iii)] boundary conditions on the gauge functions 
associated with the softly-broken gauge symmetries, \emph{viz.}
\be 
\l|_{\partial M}\qquad\mbox{for \quad $\l\in \;\mathfrak g_1/\mathfrak l$}\quad ;
\ee
and 
\item[(iv)] the boundary conditions \eq{BCUV} on the Lagrange multipliers.
\end{itemize}

\subsubsection{Duality extended spectral flow with Lagrange multipliers} 

The equations of motion ${\cal R}^i\approx 0$ of the extended Lagrangian system 
$Z^i=(A,B;U,V)$ with bilinear $P$ and $Q$ structures (\emph{i.e.} linear 
${\cal F}$ and $\widetilde{\cal F}$ functions) 
are compatible with the extended flow equations $L_1  A\approx0\approx L_1  B$ (or equivalently $L_1  B'\approx0$) and 
\be L_1 U~\approx~\mu_1  V'\star (\partial_\nu  A)-\mu_2 (\partial_\nu  A)\star  V'\ ,\qquad L_1 V'~\approx~  \mu_1  V'\star (\partial_\nu  B')+\mu_2 (\partial_\nu  B')\star  V'\ ,\ee
with flow operator $L_1$ given by \eq{flowL} and the redefinition 
\be  B~=~ \nu{\bf 1}+g B'\ ,\qquad  V~=~g  V'\ ,\qquad \nu,g~\in~\Comp\ .\ee 
We have not found any generalization of the spectral flow to the Lagrangian systems with higher-order $P$- or $Q$-structures (\emph{i.e.} nonlinear 
${\cal F}$ or $\widetilde{\cal F}$ functions).

\subsubsection{Consistent truncations off shell}

Reality conditions can be imposed off shell by requiring the action to be either 
real or purely imaginary, \emph{viz.}
\be 
(S^{\rm cl}_{\rm bulk})^\dagger~=~\e_S S^{\rm cl}_{\rm bulk}\ ,
\ee
leading to the following reality conditions on the Lagrange multipliers and the 
function $\widetilde{\cal F}$ appearing in the generalized $P$-structure:
\be ( U_{[p]})^\dagger ~=~\e_S (\e_{\Real})^{n+\frac p2}  U_{[p]}\ ,
\qquad ( V_{[p]})^\dagger~=~ -\e_S (\e_{\Real})^{n+\frac{p+1}2} V_{[p]} ,
\ee\be 
(\widetilde{\cal F}_0(\l))^\dagger~=~-\e_{\Real} 
\widetilde{\cal F}_0(\e_S(\e_{\Real})^n  \l^\dagger)\ ,\qquad
\left(\widetilde{\cal F}_I(\l)\right)^\dagger ~=~ 
\widetilde{\cal F}_{\bar I}(\e_S(\e_\Real)^n \lambda^\dagger)\ , \ee
\be \left(\widetilde{\cal F}_{I\bar J}(\lambda)\right)^\dagger~=~ 
\e_\Real \widetilde{\cal F}_{J\bar I}(\e_S(\e_\Real)^n \l^\dagger)\ .
\ee
{}From ${\rm Tr}[\pi_k(\cdot)]={\rm Tr}[\cdot]$ it follows that in the case of 
$\pi_k$-projection then the $k$-parities must be correlated as follows: 
\be \pi_k( U_{[p]})~=~-\e_k^{n+\frac p2}  U_{[p]}\ ,\qquad 
\pi_k( V_{[p]})~=~\e_k^{n+\frac{p+1}2}  V_{[p]}\ ,
\ee\be \widetilde{\cal F}_0(-(\e_k)^n \l)~=~\e_k {\widetilde{\cal F}}_0(\l)\ ,
\qquad \widetilde{\cal F}_I(-(\e_k)^n \l)~=~(-1)^{I+1}\widetilde{\cal F}_I(\l)
\ ,\ee
\be \widetilde{\cal F}_{I\bar J}(-(\e_k)^n\l)~=~\e_k 
(-1)^{I+\bar J}\widetilde{\cal F}_{I\bar J}(\l)\ .
\ee
To fix spacetime parity one may impose ($\e,\tilde\e=\pm1$)
\be P(U_{[p]})~=~\e(\e\tilde\e)^{n+\frac p2} U_{[p]}\ ,\qquad 
P(V_{[p]})~=~(\e\tilde \e)^{n+\frac{p+1}2} V_{[p]}\ ,\ee
\be 
\widetilde{\cal F}_0(\e(\e\tilde\e)^n\l)~=~\e\tilde\e{\widetilde{\cal F}}_0(\l)\ ,\qquad 
\widetilde{\cal F}_{\bar I}(\l)~=~{\widetilde{\cal F}}_I(\e(\e\tilde\e)^n\l)\ ,\qquad 
\widetilde{\cal F}_{I\bar J}(\l)~=~\e\tilde\e \widetilde{\cal F}_{J\bar I}
(\e(\e\tilde\e)^n\l)\ .
\ee
Finally, assuming $ {\rm Tr}[\tau(\cdot)]={\rm Tr}[\cdot]$, the projection to 
the minimal bosonic model takes the form
\be \tau( U_{[p]})~=~(-1)^{n+\frac p2}  U_{[p]}\ ,\qquad 
\tau( V_{[p]})~=~(-1)^{n+\frac{p-1}2}  V_{[p]}\ ,\ee\be 
\widetilde{\cal F}_0((-1)^n\l)~=~\widetilde{\cal F}_0(\l)\ ,\qquad 
\widetilde{\cal F}_I((-1)^n\l)~=~\widetilde{\cal F}_I(\l)\ ,\ee
\be \widetilde{\cal F}_{I\bar J}((-1)^n\l)~=~\widetilde{\cal F}_{I\bar J}(\l)\ .
\ee

\subsection{Even-dimensional bulk ($ p\in 2\mathbb N+1$)}

In the case of an even-dimensional bulk, say of dimension 
$ \hat{p}+1=2n$, 
one has the action
\be S^{\rm cl}_{\rm bulk}[ A, B; S, T]~=~ \int_M  {\rm{Tr}}\left[  
S \star  D  B
+ T\star ( F+{\cal F})+{\cal W}( S;  J^I, J^{\bar I},J^{ I\bar J})\star  
T\right]\ ,
\ee
where ${\cal W}$ is an interaction $\star$-function obeying
\be 
{\cal W}(-\l)~=~{\cal W}(\l)\ ,\qquad {\cal W}(0)~=~0\ ,
\ee
and the form degrees are assigned as follows: 
\be  A~=~\sum_{m=1,3,\dots,  \hat{p}}  A_{[m]}\ ,\qquad  B~=~
\sum_{m=0,2,\dots,  \hat{p}-1} B_{[m]}\ ,\ee\be
 S~=~\sum_{m=1,3,\dots,  \hat{p}}  S_{[m]}\ ,\qquad  
T~=~\sum_{m=0,2,\dots,  p-1} T_{[m]}\ .\ee
The variational principle yields the generalized curvatures
\be {\cal R}^{ A}~=~ F+ {\cal U} +
{\cal W}(S)\ ,\qquad {\cal R}^{ B}~=~ 
D B -( T\partial_{ S})\star {\cal W}( S)\ ,\ee
\be {\cal R}^{ S}~=~ D S+ (T\partial_B)\star {\cal F}\ ,
\qquad {\cal R}^{ T}~=~ D T+[ S, B]_\star\ .\ee
The action is gauge invariant and the equations of motion are integrable in the 
case of
\be \mbox{bilinear $Q$-structure}~:~ {\cal F}~=~ J\star B\ ,
\ee
for which the integrability of ${\cal R}^{ T}$ follows using the identity 
\be 
\{  S, ( T\partial_{ S})\star {\cal W}\}_\star~\equiv~[ T,{\cal W}]_\star\ ,
\ee
that holds for general even $\star$-functions ${\cal W}$.  The Cartan gauge 
transformations off shell are given by the on-shell transformations. 

\section{Discussions}\label{sec:discussions}

Let us summarize our results, speculate on future directions and finally 
conclude by trying to place our work and ideas into the more general state of 
affairs.

\subsection{Summary}

In this paper we presented an action principle for a duality extended version of 
Vasiliev's equations for interacting higher spin gauge fields (including 
gravity) in four dimensions.

The duality extended version consists of differential forms of degrees 
$p\in \{0,1,2,\dots\}$ forming two master fields $B=B_{[0]}+B_{[2]}+\cdots$ and 
$A=A_{[1]}+A_{[3]}+\cdots$, and their Lagrange multipliers which are 
differential forms of dual form degrees of degrees $\hat p-p$ where $\hat p+1$ 
is the dimension of the base manifold (including the twistor $Z$-space).
The initial and boundary data associated with the Lagrange multipliers are 
removed by means of boundary conditions compatible with the variational 
principle.
As a result, the Lagrange multipliers can be set equal to zero on shell, 
leaving $A$ and $B$ subject to the unfolded equations of motion 
${\rm d}A+A\star A+J\star B\approx 0$ and 
${\rm d}B+A\star B-B\star A\approx 0$ where 
$J=J_{[2]}+J_{[4]}$ is a closed and central element. This system contains 
Vasiliev's original equations in degrees zero and one, \emph{viz.} 
$ {\rm d} A_{[1]}+A_{[1]}\star A_{[1]}+J_{[2]}\star B_{[0]}\approx 0$ and 
${\rm d}B_{[0]}+A_{[1]}\star B_{[0]}-B_{[0]}\star A_{[1]}\approx 0$.

An important point that remains to be established is whether the coupling 
$J_{[4]}\star B$ is nontrivial in the sense that it cannot be redefined away. 
In Vasiliev's original system, the coupling $J^2_{[2]}\star B$ (and its 
hermitian conjugate) is nontrivial; it is indeed this term that reproduces the 
nontrivial interactions in the second order in curvature in the effective 
unfolded equations of motion in the perturbative expansion around a 
non-degenerate vierbein \cite{Vasiliev:1989yr}. 
The reason $J^2_{[2]}\star B$ is nontrivial is that the central term 
$J^2_{[2]}$ contains the inner Kleinian $\kappa$ (that becomes a Dirac delta 
function in the Weyl order of the $(Y,Z)$-oscillator algebra). 
We note that also $J_{[4]}$ contains such ``singular'' elements, namely 
$J^{1\bar{2}}_{[4]}\star B$ (and its hermitian conjugate) and 
$J^{\bar{1}\bar{2}}_{[4]}\star B$.

The duality-extended $(A,B)$-system is perturbatively equivalent to Vasiliev's 
original $(A_{[1]},B_{[0]})$-system:
\begin{itemize}
\item[i)] both systems share the same Weyl zero-form $B_{[0]}$; this master 
field contains the initial data associated with the Weyl curvature tensors, which 
contain one-particle states and other local deformations of the system such as 
for example the massive parameters of the black-hole solution 
of~\cite{Didenko:2009td}.
\item[ii)] the master fields with positive form degree (including $A_{[1]}$) 
bring gauge functions on shell. In topologically broken phases, the boundary 
values of gauge functions associated with topologically broken gauge symmetries 
may contribute to observables; see Appendix \ref{app:global}. 
Thus the original and duality-extended systems share the same observable gauge 
functions in the unbroken phase (where no gauge functions are observable) 
and in broken phases where projections of $A_{[1]}$ are broken (such as for 
example the $\pi$-odd projection containing the ordinary vierbein).
\end{itemize}
We wish to stress, however, that if one has an exact solution to the 
duality-extended $(A,B)$-system, then it by construction contains an exact 
solution to 
the original system. As known from \cite{Iazeolla:2007wt}, there exist exact 
solutions of the original system for which the connections exhibit critical 
behaviors for finite amplitudes of $B_{[0]}$ (as can be described invariantly 
using zero-form invariants). Thus it is not clear whether a given exact solution 
to the original system can be uplifted to the duality-extended system, as 
new critical phenomena may arise in potentials in the duality-extended sector.

We also wish to stress that the action principle involves an integration over a 
base manifold given by the product of an ordinary commuting base manifold 
(containing four-dimensional spacetime) and the non-commutative twistor 
$Z$-space. The Lagrangian also contains an additional integration over the 
internal twistor $Y$-space --- which one may think of as contracting indices 
related to various representations of an internal higher-spin Lie algebra.

In this sense, if one was to take our action principle seriously as a starting 
point for quantizing higher-spin gravity, one would have to address the issue of 
boundary conditions on the internal connection $(A_\a,A_{\ad})$ in $Z$-space. 
In the standard perturbative expansion in the Weyl zero-form $B_{[0]}$, it is 
usually assumed that $(A_\a,A_{\ad})$ is pure gauge in the limit where $B_{[0]}$ 
vanishes. However, as found in \cite{Iazeolla:2007wt}, there are 
``topologically nontrivial'' exact solutions based on projectors in which 
$(A_\a,A_{\ad})$ remains nontrivial for vanishing $B_{[0]}$, whose physical 
meaning remains to be understood better.

\subsection{Outlook: AKSZ-BV quantum action and unfolded quantum field theory}

The action principle proposed in this paper is an example of a generalized 
Hamiltonian action principle for an associative free differential algebra on a 
noncommutative base manifold. More generally, as far as the off-shell 
formulation of free differential algebras is concerned, one may think of three 
different levels of complexity depending on whether the algebra is associative 
and commutative, or associative and non-commutative, or of strongly homotopy 
associative type. In the commutative case, the BV quantum action is of the 
AKSZ-BV type and it has been proposed that the perturbative quantization (with 
suitable boundary conditions on Lagrange multipliers) yields master theories of 
the homotopy type (with $\ell$-ary products arising via terms in the Hamiltonian 
that are of $\ell$-th order in the Lagrange multipliers).

In our case, there exists a quantum action of AKSZ-BV type which we shall 
present elsewhere. Moreover, the classical $(A,B;U,V)$ system extends naturally 
to the strongly homotopy associative case  and there are indications that its 
completion off shell leads to an AKSZ-BV-like quantum action (within a suitable 
Noether procedure). It is thus tempting to speculate that there exist quantum 
theories based on layers on ``$n$-quantized'' unfolded quantum field theories 
such that each layer is the master theory of the layer below with radiative 
corrections interpreted as a topological sum, giving rise to third-quantization.

Pursuing these ideas, one is led to attempt to identify Vasiliev's equations as 
the master equations for an underlying first-quantized topological open string: 
the system on the commutative manifold appears related to an underlying A-model; 
and the system on the noncommutative twistor space appears related 
to a B-model \cite{Engquist:2005yt}.
More generally, one may deform the bulk action with various topological vertex 
operators inserted on finite-dimensional sub-manifolds:
these are gauge-invariant functionals whose variations vanish on shell (so that 
the standard first-order action is an example of such a deformation) and
whose values on shell can be interpreted as 
amplitudes \cite{Park:2000au,Hofman:2002rv,Hofman:2002jz}.
There are many such deformations, each of which one may seek to relate to an 
underlying first-quantized dual, such as for example the holographic dual in 
three dimensions for which one may propose a topological vertex operator that is 
a four-form \cite{wipGeometry}.

The perturbations of the bulk action by various operators also provides a 
systematic approach to symmetry breaking mechanisms: for example,
one has topological mechanisms (homotopy phases), spontaneous mechanisms 
(classical solutions) and dynamical mechanisms (radiative corrections).

More radically, one may go so far as to elevate the aforementioned layered 
structure of unfolded quantum field theories into a \emph{quantum gauge 
principle}, \emph{i.e.} a set of mathematical rules that are nontrivial in the 
sense that they are meant to hold for any \emph{physical} (quantum) system. In 
particular, the Cartan integrable free differential algebra of the $n$th layer, 
with its exterior derivative $d$ (on a base manifold) and $Q$-structure (in a 
target space), should arise from the BRST operator of the $(n-1)$-quantized 
system (subject to radiative corrections but with trivial topology as the 
topological sum of the $(n-1)$st layer should correspond to the radiative 
corrections of the $n$th layer). In other words, the quantum gauge principle is 
meant to contain Cartan's version of Weyl's classical gauge principle.

In other words, the idea is that generic quantum system should \emph{not} abide 
by the quantum gauge principle making it nontrivial. We believe, however, that 
the Vasiliev system is a candidate for (a massless sector of) a system 
compatible with the quantum gauge principle.

\subsection{Conclusions}

As far as four-dimensional higher-spin gravities are concerned, the only fully 
nonlinear models that are known up to this date are those that have been 
obtained within Vasiliev's formalism. 
Vasiliev's formalism provides a general framework for higher-spin gravities 
based on free differential algebras on noncommutative manifolds taking their 
values in internal associative (super)algebras.

All models arising within this framework are based on one and the same universal 
equation of motion; different models arise by choosing different base manifolds 
and associative algebras.
In this sense, all models arising within Vasiliev's framework can be viewed as 
various Yang--Mills and supersymmetric extensions of a basic minimal bosonic 
model consisting perturbatively of a scalar field, a metric and a tower of 
Fronsdal tensors of ranks $\{4,6,\dots\}\,$.

Strictly speaking, these perturbative formulations arise only under a set of 
extra assumptions (on boundary conditions in twistor spaces); whether the 
resulting perturbative models exhaust all mathematical possibilities within the 
perturbative Fronsdal programme is an open problem though there are uniqueness 
theorems to low orders.

Remarkably, notwithstanding its somewhat peculiar features in comparison to the 
more traditional approach to lower-spin gravities, the perturbative expansions 
of Vasiliev's equations around its anti-de Sitter vacuum appear paradigmatic as 
far as holography is concerned, that is, it reproduces the simplest possible 
candidates for holographic duals of higher-spin gravities 
\cite{Sundborg:2000wp,Sezgin:2001zs,Sezgin:2002rt,Klebanov:2002ja}; 
see for example the recent works in 
\cite{Giombi:2009wh,Giombi:2010vg,Koch:2010cy,Douglas:2010rc}.

Vasiliev's equations admit, however, exact solutions that involve moduli that 
are not visible in the perturbative Fronsdal Programme (for example solutions 
activating the internal connections in twistor space but not the Weyl tensors). 
The formalism also admits extensions by differential forms whose exterior 
derivatives vanish identically in the linearized approximation which one may 
think of as analogs of the three-dimensional gauge fields\footnote{These forms 
appear in the $k$-independent part of $B_{[0]}$ and the $k$-linear part of 
$A_{[1]}$.}.

Taken altogether, the state of affairs motivates a more careful examination of 
whether the full field content of Vasiliev's unfolded formalism should be 
treated as the actual fundamental field content. In Vasiliev's system, Fronsdal's 
equations appear in a precise perturbative sector and most likely the complete 
theory requires to consider the twistor $Z$-space on an equal footing with 
spacetime.
In this approach, the aim becomes to include all unfolded variables 
(differential forms) into the action principle, which leads more or less 
directly to the type of generalized Hamiltonian bulk actions considered in this 
paper and in fact already considered in \cite{Vasiliev:1988sa}, 
albeit in its simpler version without any Poisson structure. 

These action principles lend themselves naturally to the BRST treatment leading 
to generalized AKSZ-BV models, which is the stage at which we are now.
The resulting open problem is how to connect back to the perturbative 
quantization scheme within the Fronsdal Programme with its clear physical 
interpretation.
To this end it is natural to examine various perturbations of the bulk action, 
which we leave for future studies.

{\large \bf Note added}: The results in this paper were partly presented by 
P.S. at the IVth International Sakharov Conference on Physics, 18-23 May 2009, Lebedev Institute (Moscow),
and at the International workshop on Gauge Theories, Supersymmetry, and Mathematical Physics, 6-10 April 2010, Lyon, France.

{\large \bf Acknowledgement}: We acknowledge C. Iazeolla and A. Sagnotti for 
collaborations at the earliest stages of this project. We have also benefitted 
from interactions with N. Colombo, S. Lyakhovich and E. Sezgin. 
We thank I. Bandos, G. Barnich, M. Grigoriev, E. Skvortsov, D. Sorokin 
and M. Vasiliev for discussions. 
N. B. acknowledges F. Buisseret and E. Skvortsov for encouragements. 
P. S. acknowledges Ph. Spindel for encouragement. We are both grateful to 
Scuola Normale Superiore (Pisa) for support at the early stages of the project.
This work has been partly supported by the ÒActions de Recherche Concert\'eesÓ of the ÒDirection de la Recherche 
scientifique -- Communaut\'e Fran${\c c}$aise de BelgiqueÓ.


\begin{appendix}

\section{Free differential algebras on non-commutative base manifolds}\label{app:global}

Vasiliev's on-shell formulation of higher-spin gravity makes use of a version of unfolded dynamics that is based on associative free differential algebras with central and closed terms. Such an algebra encodes the following key structures:
$$(\mathfrak B,\mathfrak A,\star,{\rm d};\mathfrak J;{\cal I},\overrightarrow{\cal Q};\mathfrak t)\ ,$$
and it describes the moduli space ${\cal M}_{\mathfrak t}$ of $\mathfrak A$-valued sections $\{Z^i\}_{i\in{\cal I}}$ over a noncommutative base manifold $\mathfrak B$, subject to universally Cartan integrable flatness conditions on generalized curvatures
\be {\cal R}^i~:=~{\rm d}Z^i+{\cal Q}^i(Z,J)~\approx~0\ ,\qquad i~\in~{\cal I}\ ,\label{calR}\ee
and defined modulo unbroken Cartan gauge transformations generated by $\mathfrak t$, a subalgebra of the Cartan gauge algebra $\mathfrak g\,$.

The $\{Z^i\}$ are the fundamental (classical) fields of the unfolded system; we refer to $Z^i$ as the master field of flavor $i\,$. 
The master fields are differential forms in degrees $p_i\equiv {\rm deg}(Z^i)\in \mathbb N$ (including zero-forms). They can be acted upon with the exterior derivative 
${\rm d}$ and composed using the associative noncommutative product $\star\equiv \star\wedge$ combining the product on $\mathfrak A$ and the composition of differential forms on $\mathfrak B$ (represented by symbols). The following rules apply:
\be {\rm deg}(Z^i\star Z^j)~=~{\rm deg}(Z^i)+{\rm deg}(Z^j)\ ,\qquad 
{\rm deg}({\rm d})~=~1\ ,\ee\be
{\rm d}(Z^i\star Z^j)-({\rm d}Z^i)\star Z^j-(-1)^{{\rm deg}(Z^i)} Z^i\star(dZ^j)\equiv 0\ ,\ee\be(Z^i\star Z^j)\star Z^k-Z^i\star (Z^j\star Z^k)~\equiv~0\ .\ee
Locally, in the coordinate charts $\mathfrak B_\xi\subset \mathfrak B$, labelled here by an additional chart index, the sections have local representatives
\be Z^i_\xi~\in~\Omega(\mathfrak B_\xi)\otimes \mathfrak A\ .\ee

The structure functions ${\cal Q}^i(Z,J)$ in \eq{calR} are given by $\star$-power expansions in $Z^i$ and an additional set $\{J^I\}$ of globally defined elements that are central and closed, \emph{viz.}
\be J^I~\in~\Omega(\mathfrak B)\otimes \mathfrak A\ ,\qquad 
{\rm d}J^I~\equiv ~0\ ,\qquad J^I\star Z^i-Z^i\star J^I~\equiv~0\ ,\ee
hence generating a closed and central subalgebra
\be \mathfrak J~\subset~ \Omega(\mathfrak B)\otimes \mathfrak A\ .\ee
The structure functions can thus be presented as
\be {\cal Q}^i~=~\sum_n {\cal Q}^i_{j_1,\dots,j_n}(J^I) \star Z^{j_1}\star\cdots\star Z^{j_n}\ee
with coefficients ${\cal Q}^i_{j_1,\dots,j_n}(J^I)\in \mathfrak J$ that need not be graded symmetric in their lower flavor indices (due to the non-commutativity of $\star$). The universal Cartan integrability of \eq{calR} is tantamount to compatibility with 
${\rm d}^2\equiv0$ on base manifolds $\mathfrak B$ of arbitrary dimension. Using the notation for $\star$-vector fields (see Appendix \ref{app:c}), this amounts to that
\be \overrightarrow {\cal Q}\star \overrightarrow {\cal Q}~\equiv~ 0\ ,\qquad \overrightarrow {\cal Q}~:=~{\cal Q}^i(Z^j,J^I)\partial_i\ ,\ee
or equivalently, that the coefficients obey
\be \sum_{n_1+n_2=n-1}\sum_{m=1}^{n_1}{\cal Q}^i_{j_1,\dots,j_{m-1},k,j_{m},\dots,j_{n_1-1}}(J^I) \star {\cal Q}^k_{j_{n_1},\dots,j_n}(J^I)~\equiv~0\ ,\ee
where the flavor indices $j_1,\dots,j_n$ are not subject to any graded symmetry.

The universal Cartan integrability implies that the constraint surface remains invariant under the Cartan gauge transformations
\be \delta_\varepsilon Z^i~\equiv~\overrightarrow {\cal T}_{\varepsilon,{\rm d}\varepsilon;Z} \star Z^i~:=~
{\rm d}\varepsilon^i-\overrightarrow\varepsilon \star {\cal Q}^i\ ,\qquad \overrightarrow\varepsilon~:=~\varepsilon^i\partial_i\ ,\ee
which are linear in gauge parameters $\varepsilon^i$ and in general nonlinear in $Z^i$. These transformations form a soft gauge algebra $\mathfrak g$ that exponentiates into generalized (or soft) group elements
\be {\cal G}_{\l,Z}~:=~\exp_\star \overrightarrow {\cal T}_{\l,{\rm d}\l;Z}\label{softG}\ee
generated by (finite) gauge functions $\l^i\,$. 
The space ${\cal M}_\xi$ of locally defined solutions to 
${\cal R}^i_\xi:={\rm d}Z^i_\xi+{\cal Q}^i(Z^j_\xi,J^I)\approx0$ is given formally by Cartan gauge orbits, \emph{viz.}
\be {\cal M}_\xi~=~\{{\cal G}_{\l,Z} \star Z^i~:~\l=\l_\xi\,, \,Z^i=Z^i_{C_\xi}\}\ ,\ee
where $\l^i_\xi$ and $Z^i_{C_\xi}$ are locally-defined gauge functions and reference solutions, respectively; the reference solution obeys
i) the constraints ${\rm d}Z^i_{C_\xi}+{\cal Q}^i(Z_{C_\xi},J)\approx 0$;  ii) an initial datum $(Z^i_{C_\xi}|_{[0]})|_{p_\xi}=C^i_\xi$ where $p_\xi\in\mathfrak B_\xi$ is a base point and $(\cdot)|_{[0]}$ denotes the projection to zero form degree; and iii) a physical gauge condition (to select a well-defined particular solution and avoid over-representation). Interestingly enough, the unfolded formulation of higher-spin gravities appears amenable to the implementation of the above form of Cartan integrability at least in sub-sectors of the theory.

The moduli space ${\cal M}_{\mathfrak l}$ is obtained by first gluing together 
locally-defined modules ${\cal M}_\xi$ by means of transition functions valued in the unbroken gauge algebra $\mathfrak l\subseteq\mathfrak g$, \emph{viz.}
\be {\cal M}_\xi~\cong~ {\cal G}_\xi^{\xi'}\star {\cal M}_{\xi'}\ ,\qquad  {\cal G}_\xi^{\xi'}~:=~\exp_\star \overrightarrow {\cal T}_{t_\xi^{\xi'},{\rm d}t_\xi^{\xi'};Z^i_{\xi'}}\ ,\qquad t_\xi^{\xi'}~\in~\mathfrak l\ ,\ee
where the parameters are defined on (cylinders homotopic to) the overlaps $\mathfrak B_\xi\cap \mathfrak B_{\xi'}$ (we are assuming that $\mathfrak B=\bigcup_\xi \mathfrak B_\xi$). The gluing compatibility implies that
\be Z^i_{C_\xi}~=~Z^i_C\qquad\mbox{for all $\xi$}\ ,\ee
where thus $C$ is (gauge non-invariant) constant of motion, and that
\be {\cal G}_\xi^{\xi'}~=~ {\cal G}_\xi\star ({\cal G}_{\xi'})^{-1}\ ,\label{Gxixiprime}\ee
which is a nontrivial gluing condition on the gauge functions.
The coordinates on ${\cal M}_{\mathfrak l}$ are gauge-invariant and intrinsically defined observables ${\cal O}_{\mathfrak l}$, that is, functionals of the master fields constructed out of local functionals that are manifestly $\mathfrak l$-invariant off shell and intrinsically defined, \emph{i.e.} independent under any particular choices of local data on the base manifold and hence manifestly diffeomorphism invariant (consequently non-local). The manifest $\mathfrak l$-invariance implies that 
${\cal G}_\xi^{\xi'}\sim {\cal U}_\xi \star {\cal G}_\xi^{\xi'}\star ({\cal U}_{\xi'})^{-1}$ where ${\cal U}_\xi$ is generated by $\mathfrak l\,$. 
Thus, in view of \eq{Gxixiprime}, one has that
\be  {\cal G}_\xi~\sim~{\cal U}_\xi \star {\cal G}_\xi\qquad\mbox{where ${\cal U}_\xi$ is generated by $\mathfrak l$,}\ee
that is, the gauge functions in ${\cal M}_{\mathfrak l}$ can be taken to be valued in the coset $\mathfrak g/\mathfrak l$.

For example, one may consider homotopy charges given by integrals
\be {\cal O}~:=~\oint_{\Sigma'}(\omega^R+k^R)\ ,\qquad \Sigma'~\in~[\Sigma]\ee
over nontrivial $p_R$-cycles $[\Sigma]$ of $p_R$-forms $\omega^R[Z,J]$ and $k^R[Z,J]$ that are manifestly $\mathfrak l$-invariant, \emph{i.e.}
\be \delta_\varepsilon (\o^R,k^R)~\equiv~0\ ,\qquad \varepsilon\in \mathfrak l\ ,\ee
and defined by the equivariant cohomology system
\be {\rm d}\omega^R+f^R(\omega)~\approx0\ ,\qquad f^R(\o)|_{\Sigma_{\rm cyl}}~\approx~ 
{\rm d}k^R|_{\Sigma_{\rm cyl}}\ ,\ee
where $\Sigma_{\rm cyl}$ is a cylinder of finite thickness containing $\Sigma$; the homotopy invariance of de Rham cohomology classes then implies that $H^{p_R+1}(\Sigma_{\rm cyl})=0$ so that $f^R|_{\Sigma_{\rm cyl}}$ must be exact, that is, given by the exterior derivative of some $p_R$-form $k^R$ that is globally defined on $\Sigma$ (and hence gauge invariant). Thus the integral over $\Sigma$, which must necessarily be split into several charts, say $\{\Sigma_\xi\}$, makes sense and is independent of the choice of $\Sigma'$. A variation $\delta_\varepsilon \l^i=\varepsilon^i$ in the gauge functions thus induces a change in $(\o^R+k^R)|_{\Sigma_\xi}$ given by 
\be \delta_{\varepsilon} (\o^R+k^R)|_{\Sigma_\xi}~=~{\rm d} X_\xi(\varepsilon_\xi)\ ,\ee
where $X_\xi(\varepsilon_\xi)$ is a linear functional in $\varepsilon^i_\xi$. By the $\mathfrak l$-invariance, one has that $X_\xi(\varepsilon_\xi)$ is invariant under 
$\mathfrak l$-transformations that act simultaneously on $Z^i$ and the gauge parameter (\emph{c.f.} the BRST treatment where the gauge parameter is promoted into a ghost). It follows that
\be \delta_{\varepsilon}{\cal O}_{\mathfrak l}~=~\sum_{\xi} \oint_{\partial\Sigma_{\xi}} X_\xi(\varepsilon_\xi)\ ,\ee
which can be split into contributions from chart boundaries in the interior of $\mathfrak B$ and from true boundaries of $\mathfrak B$. The former must cancel identically if one assumes that the choice of where to cut the interior of $\mathfrak B$ into charts should not be of no importance. Taking into account the signs coming from orientation, this is a consequence of the fact that $\{\l^i\}$ forms a globally defined section (of the soft $\mathfrak l$-bundle) as stated in \eq{Gxixiprime}. One thus has
\be \delta_{\varepsilon}{\cal O}_{\mathfrak l}~=~\sum_\xi \oint_{\partial\mathfrak B\cap \partial\Sigma_\xi} X_\xi(\varepsilon_\xi)\ ,\ee
that is the only physical dependence of the gauge functions enters via their boundary values, which one may view as an unfolded version of the holographic principle.

\section{Duality extension}\label{app:extension}

We consider an associative free differential algebra consisting of master fields $Z^i$ and structure coefficients ${\cal Q}^i_{j_1,\dots,j_n}(J^I)$ of fixed degrees, say ${\rm deg}(Z^i)= p_i \in \mathbb N$ and ${\rm deg}({\cal Q}^i_{j_1,\dots,j_n})= p^i_{j_1\dots j_n}\in 2\mathbb N$. This system can always be duality extended (without adding any new local degrees of freedom) by i) replacing $Z^i$ by $\widehat Z^i:=\sum_k Z^i_{[p_i+2k]}$; and ii) exploiting field redefinitions to introduce coupling constants $g_{[0]}$ and then replace these by $\widehat g(J^I):=\sum_k g_{[2k]}$. It follows that the extended system $\{\widehat Z^i,\widehat g\}$ contains the original system $\{Z^i_{[p_i]},g_{[0]}\}$ as a consistent subsystem, though the added master fields $Z^i_{[p_i+2k]}$ with $k>0$ cannot in general be set equal to zero, since they are sourced from $\{Z^i_{[p_i]}\}$ via terms involving the new couplings $g_{[2k]}$ with $k>0$.

One may refer to the duality extension as non-trivial if the central elements cannot be removed by redefining the master fields; we are not aware of any general condition that guarantees non-triviality.

\section{Further details: $\star$-vector fields and Cartan integrability}\label{app:c}

In this Appendix we go into the technical details of $\star$-functions, $\star$-vector fields and Cartan integrability that were introduced in Appendix \ref{app:global}. Let us first recall the general idea of a free differential algebra on a non-commutative base manifold $\mathfrak B$ consisting of graded associative algebras ${\mathfrak{R}}_{\xi}$ generated by sets $\{Z^i_\xi\}_{i\in{\cal S}}$ of locally-defined differential forms subject to generalized curvature constraints
\begin{eqnarray}
 {\cal R}^i_\xi~:=~{\rm d} Z^i_\xi+{\cal Q}^i(Z_\xi,J)~\approx~0\ ,
\end{eqnarray}
where $\overrightarrow{\cal Q}:={\cal Q}^i\,\partial_i$ is a composite $\star$-vector field
of total degree one subject to the Cartan integrability condition
\begin{eqnarray}
\overrightarrow{\cal Q}\star {\cal Q}^i~\equiv~0\ .\label{CIcond}
\end{eqnarray}
Here we use the following notation and conventions:
\begin{enumerate}
 \item[(i)] $\xi$ labels charts $\mathfrak B_\xi\subset \mathfrak B$ with 
 coordinates $\Xi^M_\xi$ of
 degree zero and differentials ${\rm d}\Xi^M_\xi$ of degree one generating
 $\mathbb N$-graded associative $\star$-product algebras
 \begin{eqnarray}
 \Omega_\xi~\equiv~ {\rm Env}[\Xi^I_\xi,{\rm d}\Xi^I_\xi]
 \end{eqnarray}
 modulo the graded $\star$-commutators
 \begin{eqnarray}
 [\Xi^M_\xi , \Xi^N_\xi]_\star~=~2i\Pi^{MN}\ ,\quad [\Xi^M_\xi ,
 {\rm d}\Xi^N_\xi]_\star~=~0\ ,\quad 
 [{\rm d}\Xi^M_\xi , {\rm d}\Xi^N_\xi]_\star~=~0\ ,
 \end{eqnarray}
 where $\Pi^{MN}$ is a constant matrix (defining a canonical Poisson structure
 $\Pi=\Pi^{MN}\partial_M\otimes \partial_N$);
 \item[(ii)] the action of the exterior derivative 
 ${\rm d}={\rm d}\Xi^M_\xi\partial/\partial \Xi^M_\xi$ in $\O_\xi$ is
 defined by declaring that
 \begin{eqnarray}
 {\rm d}(\Xi^M_\xi)~=~{\rm d}\Xi^M_\xi\ ,\qquad {\rm d}(f\star g)~=~
             ({\rm d}f)\star g+(-1)^{{\rm deg}f} f\star ({\rm d}g)\ ,
 \end{eqnarray}
 for elements $f,g\in\O$ such that $f$ has fixed form degree ${\rm deg} f$\,; one has
 \begin{eqnarray}
 {\rm d}^2\equiv 0\ .
 \end{eqnarray}
 \item[(iii)] the locally-defined differential forms
 $Z^i_\xi \in \Omega^{[p_i]}_\xi \otimes \Theta^i\,$,
 where $\Omega^{[p_i]}_\xi$ is the subspace of $\O_\xi$ of fixed form degree
 $m_i$ and $\Theta^i$ can be either are finite-dimensional internal tensors
 (such as for example Lorentz tensors) or sectors of an internal associative algebra $\mathfrak A$;
 \item[(iv)] the graded associative $\star$-product algebra
 ${\mathfrak{R}}_\xi:={\rm Env}[Z^i_\xi]\otimes \mathfrak J$
 where $\mathfrak J$ is a space of central and $d$-closed elements (including the
 identity), \emph{i.e.} if ${\cal F}(Z^i_\xi)\in  {\mathfrak{R}}_\xi$ then
 \begin{eqnarray}
 {\cal F}~=~\sum_{n\geq 0} {\cal F}_{j_1\dots j_n}\star \,Z^{j_1}_\xi\star
 \cdots\star Z^{j_n}_\xi\ ,\qquad {\cal F}_{j_1\dots j_n}~\in~ \mathfrak J\ ;
 \end{eqnarray}
 \item[(v)] a composite $\star$-vector field $\overrightarrow{\cal X}$ is a graded inner derivation
 of ${\mathfrak{R}}$, \emph{i.e.} if ${\cal F},{\cal F}'\in {\mathfrak{R}}$ then
 \begin{eqnarray}
 \overrightarrow{\cal X}\star ({\cal F}\star {\cal F}')~=~(\overrightarrow{\cal X}\star {\cal F})\star
 {\cal F}'+(-1)^{{\rm deg}(\overrightarrow{\cal X}){\rm deg}({\cal F})}{\cal F}\star
 (\overrightarrow{\cal X}\star {\cal F}')\ , \label{Leibnitz}
 \end{eqnarray}
 provided that $\overrightarrow{\cal X}$ and ${\cal F}$ have fixed degrees.
 In components, one writes $\overrightarrow{\cal X}:={\cal X}^{ i}( Z^{ j}) \partial_{ i}$
 where
 ${\cal X}^{ i}:={\cal X}\star Z^{ i}$ (and $\partial_i\equiv \overrightarrow\partial_i$).
 The graded bracket between two composite $\star$-vector fields is defined by
 \begin{eqnarray}
 [\overrightarrow{\cal X},\overrightarrow{\cal X}']_{\star}\star {\cal F}~:=~\overrightarrow{\cal X}\star(\overrightarrow{\cal X}'\star
 {\cal F})-(-1)^{{\rm deg}(\overrightarrow{\cal X}){\rm deg}(\overrightarrow{\cal X}')}\overrightarrow{\cal X}'\star
 (\overrightarrow{\cal X}\star{\cal F})\ ,
 \end{eqnarray}
 is a degree-preserving graded Lie bracket, \emph{i.e.}
 $[\overrightarrow{\cal X},\overrightarrow{\cal X}']_{\star}$ is a graded inner derivation  obeying the graded
 Jacobi identity
 $[[\overrightarrow{\cal X},\overrightarrow{\cal X}']_{\star},\overrightarrow{\cal X}'']_\star+\mbox{graded cyclic}\equiv 0\,$.
 In  components, one has
 \begin{eqnarray}
 [\overrightarrow{\cal X},\overrightarrow{\cal X}']_{\star} ~=~ \left(\overrightarrow{\cal X}\star {\cal X}^{\prime i}
 - (-1)^{\overrightarrow{\rm deg}({\cal X}){\rm deg}(\overrightarrow{\cal X}')}\overrightarrow{\cal X}'\star
        {\cal X}^i\right)\partial_i\ .
 \end{eqnarray}
\end{enumerate}
The Cartan integrability condition (\ref{CIcond}), that can be rewritten
$[\overrightarrow{\cal Q},\overrightarrow{\cal Q}]_\star\equiv 0\,$, amounts to that
$\overrightarrow{\cal Q}$ is a nilpotent composite $\star$-vector field of degree one.
This condition ensures that the generalized curvature constraints
${\cal R}^i\approx 0$ are compatible with ${\rm d}^2\equiv 0$ without further algebraic
constraints on the generating elements $Z^i_\xi\,$.
One can also show that the nilpotency of $\overrightarrow{\cal Q}$ is separately equivalent to that
the generalized curvatures ${\cal R}^i$ obey the generalized Bianchi identities
\begin{eqnarray}
{\rm d}{\cal R}^i-\overrightarrow{\cal R}\star {\cal Q}^i~\equiv~0\ ,
\qquad{\rm{where}} \qquad \overrightarrow{\cal R}:={\cal R}^i \,\partial_i \ ,
\end{eqnarray}
and transform into each other under the following Cartan gauge transformations
\begin{eqnarray}
 \delta_{\varepsilon}Z^i
~\equiv ~{\cal T}_{\varepsilon}^i
 ~:=~  d\varepsilon^i - \overrightarrow\varepsilon \star {\cal Q}^i \ ,
\qquad{\rm{where}} \qquad \overrightarrow\varepsilon:=\varepsilon^i \,\partial_i  \
\end{eqnarray}
and where $\varepsilon^i$ is an element in $\Omega\otimes \Theta^i$ that is considered infinitesimal
and independent of $Z^i\,$, \emph{viz.}
\begin{eqnarray}
 \delta_{\varepsilon}{\cal R}^i~=~ - \overrightarrow{\cal R}\star
 \left( (\overrightarrow\varepsilon\star {\cal Q}^i)\right).
\end{eqnarray}
The closure relation reads
\begin{eqnarray}
 [\delta_{\varepsilon_1},\delta_{\varepsilon_2}] Z^i &=&
 \delta_{\varepsilon_{12}} Z^i - \overrightarrow{\cal R}\star\varepsilon_{12}^i
 \ ,
\end{eqnarray}
where the combined parameters $\varepsilon_{12}^i$'s are given by
\begin{eqnarray}
 \varepsilon_{12}^i~=~ -\frac{1}{2}\, [\overrightarrow\varepsilon_{1},\overrightarrow\varepsilon_{2}]_\star \star\,{\cal Q}^i \ .\label{closure}
\end{eqnarray}
The above results can easily be obtained upon introducing the even $\star$-vector field
\begin{eqnarray}
 \overrightarrow{\cal V}_{\varepsilon} &:=& (\overrightarrow\varepsilon \star {\cal Q}^i)\,\partial_i
\label{Vvector}
\end{eqnarray}
and using the following set of identities which are consequences of the first one:
\begin{eqnarray}
 [\overrightarrow{\cal Q},\overrightarrow{\cal Q}]_\star~\equiv ~0\quad, \qquad
 [\overrightarrow{\cal Q},\overrightarrow{\cal V}_{\varepsilon}]_\star~\equiv ~0\quad, \qquad
 [\overrightarrow{\cal V}_{\varepsilon_1},\overrightarrow{\cal V}_{\varepsilon_2}]_\star~\equiv ~ [\overrightarrow{\cal Q},\overrightarrow\varepsilon_{12} ]_\star \quad ,
\label{iden}
\end{eqnarray}
where we recall the all the commutators are graded-commutators.
\vspace*{.3cm}

As discussed above, the local representatives ${\mathfrak{R}}_\xi$ are glued together on overlaps $\mathfrak B_\xi\cup \mathfrak B_{\xi'}$ by means of the transitions $Z^i_\xi={\cal G}_\xi^{\xi'}\star Z^i_{\xi'}$ where the transition functions ${\cal G}_\xi^{\xi'}$ are soft group elements given by $\star$-exponentials of the Cartan gauge transformations as in \eq{softG}. From the Leibnitz' rule \eq{Leibnitz} it follows that these transitions are indeed isomorphisms, \emph{viz.}
\be
 {\cal G}\star {\cal F}(Z)~=~{\cal F}({\cal G}\star Z)\ ,\qquad
{\cal G}\star ({\cal F}\star {\cal F}')=({\cal G}\star {\cal F})\star({\cal G}\star{\cal F}')\ .\ee
We would like to show that, if $Z^i$
satisfies the star-product equation ${\rm d}Z^i + {\cal Q}^i (Z^j)\approx 0\,$,
then $Z^i_{\l}:=(\exp_\star[\overrightarrow {\cal T}_{\l,Z}])\star Z^i$
where $\overrightarrow{\cal T}_\l := {\rm d}\l^i\,\partial_i - \overrightarrow{\cal V}_\l\,$ [see (\ref{Vvector})]
satisfies the equation
${\rm d}Z_\l^i + {\cal Q}^i (Z_\l,J)\approx 0\,$, thereby exhibiting the
fundamental integrability of the unfolded equations in the case where the free differential
algebra $\cal A $ is endowed with a non-commutative star-product.  We recall that

\noindent \textbf{Lemma: }
The following commutation relation is true: 
$[\overrightarrow{\cal T}_\l,{\rm d}]_{\star}\approx 0 \,$,
where the weak equality means an equality on the surface
$\Sigma\equiv \{{\rm d}Z^i + {\cal Q}^i (Z,J)\}= 0\,$.

\noindent \textbf{Proof of the Lemma: }
On the surface $\Sigma\,$, the total exterior derivative ${\rm d}\approx \overrightarrow{\cal Q} - \overrightarrow\Lambda\,$, where
\begin{equation}
\overrightarrow\Lambda:={\rm d} \l^i \,\frac{\partial}{\partial\l^i}\quad .
\end{equation}
The proof is tantamount to showing that
$[\overrightarrow{\cal T}_\l , \overrightarrow{\cal Q} -\overrightarrow\Lambda]_{\star} \star Z^i = 0 =
[\overrightarrow{\cal T}_\l,\overrightarrow{\cal Q}-\overrightarrow\Lambda]_{\star}\star \l^i\,$
because then, using the fact that  $[\overrightarrow{\cal T}_\l , \overrightarrow{\cal Q} -\overrightarrow\Lambda]_{\star}$
is a $\star$-vector field,
it follows that $[\overrightarrow{\cal T}_\l , \overrightarrow{\cal Q} -\overrightarrow\Lambda]_{\star}\,\star {\cal F}(Z,\l)=0$
for an arbitrary star-product function ${\cal F}(Z,\l)\,$.
\begin{enumerate}
  \item[(a)] First of all, it is trivial to see that
  $[\overrightarrow{\cal T}_\l,\overrightarrow{\cal Q}-\overrightarrow\Lambda]_{\star}\star \l^i = 0\,$.
  Indeed, it gives $\overrightarrow{\cal T}_\l ( d \l^i)$ which
  vanishes\footnote{We consider the algebra where
  the fields $\{Z^{i}\}$ and $\{\l^i \}$ are considered as independent,
  in accordance with the $BRST$ treatment of gauge systems.}.
  \item[(b)] That $[\overrightarrow{\cal T}_\l,\overrightarrow{\cal Q}-\overrightarrow\Lambda]_{\star}\star Z^i = 0\,$
  is more difficult to show.
  For that, we write
  $${\cal Q}^i = \sum_n {\cal Q}^i_{j_1\ldots j_n}(J)\star\;Z^{j_1}\star\ldots\star
  Z^{j_n}$$ where ${\cal Q}^i_{j_1\ldots j_n}\in \mathfrak J$
  and compute
\begin{eqnarray}
\overrightarrow{\cal Q}\star({\cal T}_\l\star Z^i) &=& -\sum_{n}\sum_{\beta < \alpha =1}^n
(-1)^{j_{\beta +1}+\ldots+j_{\a - 1}}
{\cal Q}^i_{j_1\ldots j_n}\star\;Z^{j_1}\star\ldots\star {\cal Q}^{j_\beta}
\star\ldots\star\l^{j_\a}\star\ldots\star Z^{j_n}
\nonumber \\
&& - \sum_{n}\sum_{\a < \b=1}^n (-1)^{1+j_{\a}+\ldots+j_{\beta}}
{\cal Q}^i_{j_1\ldots j_n}\star\; Z^{j_1}\star\ldots\star \l^{j_i}
\star\ldots\star {\cal Q}^{j_\beta}\star\ldots\star Z^{j_n} \quad ,
\nonumber \\
\overrightarrow{\cal T}_\l\star({\cal Q}\star Z^i) &=& [{\rm d}\l^k - (\l^j\,\partial_j)
\star {\cal Q}^k]\partial_k \star {\cal Q}^i\quad,
\nonumber \\
\overrightarrow\Lambda\star({\cal T}_\l\star Z^i) &=& -\sum_{n}
\sum_{\a = 1}^n  {\cal Q}^i_{j_1\ldots j_n}\star \;Z^{j_1}\star\ldots\star 
{\rm d}\l^{j_\a} \star\ldots\star Z^{j_n}
\quad, \quad \quad
 {\cal T}_\l\star(\Lambda\star Z^i) ~=~ 0\quad .
 \nonumber
\end{eqnarray}
Regrouping all the terms, we find
\begin{eqnarray}
[\overrightarrow{\cal T}_\l,\overrightarrow{\cal Q}-\overrightarrow\Lambda]_\star\star Z^i &=& {\cal Q}^j\partial_j
\star[ (\l^k\partial_k)\star {\cal Q}^i]
 - [(\l^k \partial_k) \star {\cal Q}^j]\partial_j \star {\cal Q}^i
\end{eqnarray}
which vanishes identically due to the second identity of (\ref{iden}).
\end{enumerate}
Therefore, since $[\overrightarrow{\cal T}_\l,\overrightarrow{\cal Q} - \overrightarrow\Lambda]_{\star}$ is a star-product vector field,
it follows that $[\overrightarrow{\cal T}_\l, \overrightarrow{\cal Q} - \overrightarrow\Lambda]_{\star}\star\,{\cal F}(Z,\l)=0$
for an arbitrary
star-product function ${\cal F}(Z,\l)\,$. $\Box$
\vspace*{.2cm}

\noindent Using the above Lemma, we have that
$Z^{i}_\l :=(\exp_\star[\overrightarrow{\cal T}_\l])\star Z^{i} $
satisfies the equation
${\rm d}Z_\l^i + {\cal Q}^i (Z_\l^j)\approx 0\,$,
since
${\rm d} \,Z^{i}_\l \equiv {\rm d} \,[ (\exp_\star[\overrightarrow{\cal T}_\l]) \star Z^{i}] =
(\exp_\star[\overrightarrow{\cal T}_\l]) \star {\rm d}\,Z^{i} \approx - (\exp_\star[\overrightarrow{\cal T}_\l])
\star {\cal Q}^{i}(Z)\equiv -{\cal Q}^{i}(Z_\l)\,$.
This proves the formal Cartan integrability of the star-product
unfolded equations.
%
\section{The Vasiliev equations}\label{app:redef}
%

In the case of Vasiliev's equations, the master fields are locally-defined 
operators of the form
\be 
O_\xi(X^M_\xi,P^\xi_M,{\rm d}X^M_\xi,{\rm d}P^\xi_M; Z^{\una},{\rm d}Z^{\una};Y^{\una};e^i)\ ,
\ee
where the non-vanishing commutators among the coordinates are
\be[X^M,P_N]_\star~=~i\d^M_N\ ,\quad [Y^{\una},Y^{\unb}]_\star ~=~2i 
C^{\underline{\a\b}}\ ,\quad [Z^{\una},Z^{\unb}]_\star ~=~-2i 
C^{\underline{\a\b}}\ ,\label{starXPYZ}
\ee
with charge conjugation matrix\footnote{We raise and lower quartet and doublet 
indices using the conventions 
$\Lambda^{\una}=C^{\underline{\a\b}}\Lambda_{\unb}\,$, and 
$\l^\a=\e^{\a\b}\l_\b$ and $\l_\a=\l^\b\e_{\b\a}\,$, and we use the notation 
$\Lambda\cdot \Lambda'=\Lambda^\una \Lambda_\una\,$,  
$\l\cdot \l'=\l^\a \l'_\a$ and $\bar\l\cdot \bar\l'=\bar\l^\ad \bar\l'_\ad\,$.}
$C^{\a\b}=\e^{\a\b}$ and $C^{\ad\bd}=\e^{\ad\bd}\,$ 
and where $\{e^i\}\,$, $\,i=1,2\,$, are two outer Kleinian operators.
The operators are represented by symbols $f[O_\xi]$ obtained by going to 
specific bases for the operator algebra which one may also think of as ordering 
prescriptions\footnote{The symbols are thus defined modulo similarity 
transformations generated by inner automorphisms (related to the higher-spin 
gauge transformations) as well as changes of the order prescription, that is, 
changes of basis of the operator algebra. These types of transformations may 
have a drastic effect on the mathematical nature of the symbols, that may change 
from being a smooth or real analytic into being singular or even distributions. 
Thus, in order to extract physically meaningful information from the master 
fields, one needs to develop the notion of observables ${\cal O}$, namely 
functionals of the locally-defined master fields that are invariant under both 
gauge transformations and re-orderings. The construction of such functionals 
introduces various geometric concepts into the theory, such as flat connections, 
covariantly constant sections (going into decorated Wilson loops), equivariantly 
closed forms (used to define homotopy charges) and metrics (that yield minimal 
areas of closed cycles).}.
One may think of the symbols as functions 
$f(X,P,Z;Y;{\rm d}X,{\rm d}P,{\rm d}Z)$ (with variables 
composed using commutative juxtaposition) on a correspondence space
\begin{equation}
 \mathfrak C~=~\bigcup_\xi \mathfrak C_\xi\ ,\qquad \mathfrak C_\xi~=~
\mathfrak B_\xi\times \mathfrak Y\ ,\qquad \mathfrak B_\xi~=~\mathfrak M_\xi 
\times \mathfrak Z 
\label{correspondencespace}
\end{equation}
equipped with a suitable associative star-product operation $\star$ 
which reproduces, in the space of symbols, the composition rule for operators.
Working within a restricted class of orderings, referred to as universal 
orderings, the exterior derivative on $\mathfrak B$ is given by
\begin{equation}
{\rm d}~=~{\rm d}X^M\partial_M+{\rm d}P_M \partial^M+q\ ,\qquad q~:=~ 
{\rm d}Z^\una \partial_\una\ . 
\end{equation}
The master fields of the (duality-unextended) minimal bosonic model are an 
adjoint one-form
\begin{eqnarray}
& A~=~  W+ V\ ,&
\\
&W~=~{\rm d}X^M \, W_M(X,P,Z;Y)\,+\,{\rm d}P_M \, W^M(X,P,Z;Y)\ ,\qquad  
V~=~{\rm d}Z^\una\,  V_\una(X,P,Z;Y)\ ,\quad &
\end{eqnarray}
and a twisted-adjoint zero-form
\begin{equation}
 \Phi~=~  \Phi(X,P,Z;Y)\ ;
\end{equation}
these fields obey the following projection and reality conditions\footnote{Here 
we are focusing on the models containing spacetimes with Lorentzian signature 
and negative cosmological constant; for other signatures and signs of the 
cosmological constant, see \cite{Iazeolla:2007wt}.}: 
\begin{equation}
 \tau( A, \Phi)~=~(- A,\pi( \Phi))\ ,\qquad ( A, \Phi)^\dagger~=~ 
      (- A,\pi( \Phi))\ ,\label{minboscond}
\end{equation}
where the maps $\tau$\,, $\pi$\,, $\bar\pi$ and $\dagger$ are defined by
$ {\rm d}\circ (\tau,\pi,\bar\pi,\dagger)=(\tau,\pi,\bar\pi,\dagger)\circ  {\rm d}$
and\footnote{The rule $ ( f\star  g)^\dagger= g^\dagger\star  f^\dagger$ holds 
for both real and chiral integration domain.}
\bea
\label{pz1}
\pi~(y_{\a},\yb_\ad;z_\a,\zb_\ad)&=&(-y_{\a},\yb_\ad;-z_\a,\zb_\ad)\ ,
\qquad\qquad \pi( f\star  g)~=~\pi( f)\star\pi( g)\ ,\\[5pt]
\bar\pi~(y_{\a},\yb_\ad;z_\a,\zb_\ad)&=&(y_{\a},-\yb_\ad;z_\a,-\zb_\ad)\ ,
\qquad\qquad \bar\pi( f\star  g)~=~\bar\pi( f)\star\bar\pi( g)\ ,\\[5pt]
\t~(y_{\a},\yb_\ad;z_\a,\zb_\ad)&=&(iy_{\a},i\yb_\ad;-iz_\a,-i\zb_\ad)\ ,\qquad 
\t( f\star  g)~=~(-1)^{ f g} \t( g)\star\t( f)\ ,\\[5pt]
(y_{\a},\yb_\ad;z_\a,\zb_\ad)^\dagger&=&(\yb_{\ad},y_\a;\zb_\ad,z_\a)\ ,\qquad 
\qquad\quad
( f\star  g)^\dagger~=~(-1)^{ f g}\;  g^\dagger\star  f^\dagger\ .
 \eea
The $\tau$-projection removes all terms that are associated with the unfolded 
description of spacetime fermions as well as spacetime bosons with odd spin.

The full equations of motion for the minimal bosonic model with the simplest interaction freedom amount to the statement that the full curvature 
$ F= {\rm d}A+A\star A$ is 
proportional to $\Phi\,$, \emph{viz} $F+\Phi \star  J=0\,$, via a deformed 
symplectic two-form $ J$ that is a defined globally on correspondence 
space and obeying $\tau( J)= -J = J^\dagger$ and
\begin{equation}
   {\rm d}J~=~0\ ,\qquad \left[J,  f\right]_\pi~=~ 0\ ,\label{Jtwist}
\end{equation}
for any $ f$ obeying
$\pi\bar\pi( f)= f$\,, and where we have defined 
$\left[ f, g\right]_\pi= f\star  g- g \star\pi( f)\,$. 
In the minimal model,
\begin{equation}
 J~=~ -\frac{i}{4} (b \,dz^2~{\kappa}+\bar b \,d\zb^2~{\kappab})\ , 
\end{equation}
where the chiral Klein operators
are given in the normal-ordering by
\begin{equation}
{\kappa}~  =~ \exp(iy^{\a} z_{\a})\ ,\qquad
 {\bar{\kappa}} ~=~{\kappa} ^\dagger=\exp(-i\yb^{\ad}\zb_{\ad})\ . 
\end{equation}
The quantities $\kappa$ and 
$\bar\kappa$ are the Klein operators of the chiral Heisenberg algebras generated 
by $(y_\a,z_\a)$ and $(\yb_\ad,\zb_\ad)$\,. The two-dimensional complexified 
Heisenberg algebra
$[u,v]_\star=1$ has the Klein operator 
$ \kappa = cos_\star(\pi v\star u)\,$, which 
anti-commutes with $u$ and $v$ and squares to $1\,$. 
Hence $\kappa$ remains invariant 
under the canonical $SL(2;\mathbb{C})$-symmetry that becomes manifest in Weyl 
order, where the symbol of $\kappa$ is thus proportional 
to the two-dimensional Dirac delta function. 
It follows that $(\kappa,\bar\kappa)$ is invariant under 
$SL(4;\Comp)\times \overline{SL}(4;\Comp)$\,, which is broken by $dz^2$ and 
$d\bar z^2$ down to a global $GL(2;\Comp)\times \overline{GL}(2;\Comp)$ symmetry 
of the Vasiliev system, generated by diagonal 
$SL(2;\Comp)\times \overline{SL}(2;\Comp)$ transformations and the exchange 
$(y_\a,z_\a)\leftrightarrow (iz_\a,-iy_\a)\,$. 
The latter symmetry is hidden in the formulation in terms of differentials on 
$Z$-space while it becomes manifest in the deformed-oscillator formulation.
\vspace*{.3cm}

By making use of field redefinitions 
$ \Phi\rightarrow \l F$ with $\l\in \Real$\,, $\l\neq 0\,$, 
the parameter $b$ in $J$ can be taken to obey
\begin{equation}
 |b|~=~1\ ,\qquad {\rm arg}(b)~\in~[0,\pi]\ . 
\end{equation}
The phase breaks parity except in the following two cases:
\be \mbox{Type A model (parity-even physical scalar)}~:~~b=1\ ,\ee
\be \mbox{Type B model (parity-odd physical scalar)}~:~~b=i\ .\ee
The integrability of $ F+ \Phi \star  J=0$ implies that 
$ D\Phi \star J=0$, that is, $ D\Phi=0\,$, 
where the twisted-adjoint covariant derivative 
$ D\Phi = {\rm d}\Phi + A\star \Phi - \Phi \star\pi(A)\,$. 
This constraints is integrable since 
\begin{equation}
  D^2\Phi ~=~ F\star \Phi - F\star\pi( \Phi)~=~-\Phi\star J\star \Phi 
    +  \Phi \star \pi(\Phi)\star J~=~ 0\ ,
\end{equation}
using the constraint on $ F$ and (\ref{Jtwist}).

Thus, in summary, the unfolded system describing the minimal higher-spin 
gravity 
with simplest possible interaction term, is given by\footnote{The format applies 
also to Yang--Mills 
extended or supersymmetric models; for example, see 
\cite{Sezgin:1998eh,Sezgin:2002ru,Engquist:2002vr}.}
\be  
F+\Phi\star  J~=~0\ ,\qquad  D\, \Phi~=~0\ ,\qquad  {\rm d} J=0\  ,\label{VasEoM}
\ee
\be  F~=~ {\rm d}A+A\star A\ ,\qquad  D\Phi~=~{\rm d}\Phi + \left[A,\Phi \right]_\pi\ ,\ee
and the kinematic constraints \ref{minboscond} which imply 
$\left[ A, J\right]_\pi=0=\left[\Phi, J\right]_\pi$\,.
The integrability is manifest in as much as the associativity of the 
$\star$-product in manifest.
The integrability implies the Cartan gauge transformations \footnote{These 
transformations are the canonical transformations of the $\star$-product algebra 
generated by (\ref{starXPYZ}) containing the diffeomorphisms of Lagrangian 
submanifolds of the unifold.}
\begin{equation}
\delta_{{\e}}  A~=  D  \e\ ,\qquad \delta_{{\e}} \Phi~=~ 
-\left[\e,\Phi\right]_\pi\ ,\label{gaugetransf} 
\end{equation}
for zero-form gauge parameters $\e(X,P,Z;Y)$ obeying the same kinematic 
constraints as the master one-form, \emph{i.e.} $\tau(\e)= -\e$ and 
$(\e)^\dagger =-\e$\,.
The closure of the gauge transformations reads
\be [\delta_{ \e_1},\delta_{ \e_2}]~=~\delta_{ \e_{12}}\ ,\qquad \e_{12}~=~
[ \e_1, \e_2]_\star\ ,\ee
defining the algebra ${\mathfrak{hs}}(4)$\,.
\vspace*{.3cm}

The symbols of the Kleinians are distributions on the doubled twistor space 
whose precise form depend on the choice of ordering scheme (that can thus be 
adapted to different physical problems); for example, in overall Weyl order they 
localize to Dirac delta functions (that are useful in trace calculations) while 
in overall normal order they become Gaussians (that are useful in perturbation 
theory).

The singular nature of the Kleinians implies that the source term 
$ \Phi\star  J$ cannot be absorbed into a field redefinition 
\cite{Vasiliev:1992av}. 
Moreover, upon projection of the full equations to a Lagrangian sub-manifold of 
the universal phase space, say $P_M=0$, which can be obtained in an expansion in 
the zero-form, the twistor-space source term induces nontrivial albeit 
perturbatively defined deformations of the generalized curvatures 
${\rm d}A+A\star A$ 
and $D\Phi={\rm d}\Phi+A\star \Phi-\Phi\star\pi(A)$ of the 
$\mathfrak{hs}(4)$-valued connection 
$A= A|_{Z=P=0}$ and the twisted-adjoint zero-form $\Phi= \Phi|_{Z=P=0}$. 
Upon further weak-field expansion around large spin-two gauge fields, 
\emph{i.e.} vierbein $e^{\a\ad}$ and Lorentz connection 
$(\o^{\a\b},\bar \o^{\ad\bd})$, the deformations contain the canonical 
linearized source terms for unfolded Fronsdal tensors in accordance with 
Vasiliev's central on-shell theorem.

In other words, the Vasiliev system contains a set of nontrivial equations of motion for perturbatively defined Fronsdal tensors. The full system contains, however, various other moduli that have either problematic or no description in terms of Fronsdal fields, such as classical solutions with degenerate vierbeins and topological degrees of freedom contained in the internal connection $ A_{\una}$ \cite{Iazeolla:2007wt}.

Over and above their formal Cartan integrability, the Vasiliev equations exhibit the following more powerful integrable structures:

\begin{itemize}

\item The Maurer-Cartan integrability facilitates the explicit construction of
solutions using gauge functions 
\cite{Didenko:2006zd,Iazeolla:2007wt,Didenko:2008va,Didenko:2009tc,wipCarlo} 
and the formal 
construction gauge-invariant observables \cite{wipGeometry};

\item The zero-forms $ S_\a:=z_\a-2i A_\a$ and $ S_\ad:=\zb_\ad-2i A_\ad\,$ the following generalization of Wigner's deformed oscillator algebra with local anyonic deformation parameter $\Phi$, \emph{viz.}
\begin{eqnarray}
&[ S_\a, S_\b]_\star ~=\; -2i\e_{\a\b}(1-\Phi\star\kappa)\quad ,\quad
[ S_\ad, S_\bd]_\star ~=\;-2i\e_{\ad\bd}(1-\Phi\star{\bar\kappa})\quad ,&
\nonumber \\
& [ S_\a, S_\bd]_\star ~=~0 \quad, \quad  S_\a \star \Phi+\Phi\star \pi( S_\a)\ =\ 0
\quad ,\qquad
 S_\ad \star \Phi+\Phi\star \bar\pi( S_\ad)\ =\ 0\ ,&\qquad
\end{eqnarray}
which one may also think of as describing the deformation of the symplectic structure on a submanifold of complex dimension two of the doubled twistor space (of complex dimension four).
\end{itemize}

These properties have been used to construct classical solutions in
\cite{Sezgin:2005pv,Sezgin:2005hf,Iazeolla:2007wt,Didenko:2009td,wipCarlo}, for 
perturbative calculations of the twistor-space vertices $P(W;\Phi)$ and $J(W,W;\Phi)$ in \cite{Sezgin:2002ru} 
and direct verification of the conjectured correspondence between Vasiliev's four-dimensional higher-spin gravities and three-dimensional conformal field theories \cite{Sezgin:2002rt,Klebanov:2002ja}, first in \cite{Sezgin:2003pt} 
at the level of cubic scalar self-couplings, and
recently for the complete cubics in \cite{Giombi:2009wh,Giombi:2010vg}.



\end{appendix}



\providecommand{\href}[2]{#2}\begingroup\raggedright\endgroup

\end{document}